\documentclass[preprints,article,accept,pdftex,oneauthor]{mdpi} 

\firstpage{1} 
\pubvolume{1}
\issuenum{1}
\articlenumber{0}
\pubyear{2025}

\usepackage{subfig}
\usepackage[super]{nth}
\usepackage{fix-cm} \usepackage{siunitx}
\DeclareSIUnit\knot{kn}

\def\UrlBreaks{\do\/\do-}

\Title{Maritime Activities Observed Through Open-Access Positioning Data:  Moving and Stationary Vessels in the Baltic Sea}

\Author{Moritz H\"utten \orcidA{}}

\AuthorNames{Moritz H\"utten}

\isAPAStyle{%
       \AuthorCitation{H\"utten, M.}
         }{%
        \isChicagoStyle{%
        \AuthorCitation{H\"utten, Moritz.}
        }{
        \AuthorCitation{H\"utten, M.}
        }
}

\address[1]{\vspace{-0.2cm}
GRID Inc., 3-6-7 Kita-Aoyama, Minato-ku, Tokyo 107-0061, Japan; moritz.huetten@gridpredict.co.jp}

\abstract{Understanding past and present maritime activity patterns is critical for navigation safety, environmental assessment, and commercial operations. An increasing number of services now openly provide positioning data from the Automatic Identification System (AIS) via ground-based receivers. We show that coastal vessel activity can be reconstructed from open access data with high accuracy, even with limited data quality and incomplete receiver coverage. For three months of open AIS data in the Baltic Sea from August to October 2024, we present (i) cleansing and reconstruction methods to improve the data quality, and (ii)~a journey model that converts AIS message data into vessel counts, traffic estimates, and spatially resolved vessel density at a resolution of $\sim$400 m. Vessel counts are provided, along with their uncertainties, for both moving and stationary activity. Vessel density maps also enable the identification of port locations, and we infer the most crowded and busiest coastal areas in the Baltic Sea. We find that on average, $\gtrsim$4000 vessels simultaneously operate in the Baltic Sea, and more than 300 vessels enter or leave the area each day. Our results agree within 20\% with previous studies relying on proprietary data. }


\addhighlights{yes}
\renewcommand{\addhighlights}{%
\vspace{2pt}

\noindent
\textbf{What are the main findings?}
\begin{itemize}[labelsep=2.5mm,topsep=-3pt]
\item Vessel activity in the Baltic Sea is derived from open-access positioning data.
\item Uncertainties from incomplete vessel positioning data can be precisely quantified.
\end{itemize}\vspace{3pt}
\textbf{What are the implications of the main findings?}
\begin{itemize}[labelsep=2.5mm,topsep=-3pt]
\item {Publicly available positioning data allow for effective analysis of vessel activity in coastal waters.}
\end{itemize}
}

\begin{document}


\section{Introduction}
\label{sec:intro}

Human activities at sea constitute the backbone of the world economy through resource extraction, energy production, fishing, and~the transport of most of the global cargo. As~such, maritime traffic has increased significantly in recent decades. Today, around $90\%$~of the world trade is transported through the oceans \citep{Christiansen2020}. According to the \citet{UNCTAD2024}, global seaborne trade nearly doubled between 2004 and 2024 with an average annual growth of about $3\%$, and~it is expected to grow further by about $2\%$ yearly between 2025 and 2029. With~respect to fishing, the~\citet{FAO2024} reports that over the past two decades, fishing for marine livestock remained largely stagnant. However, aquaculture production in marine areas increased by $50\%$, accounting for 2024 a third of the global marine food supply. \citet{Paolo2024} found that the number of offshore wind turbines is rapidly growing, having surpassed oil platforms globally in 2021.

This rise in maritime activities adds to the pressure on the climate \citep{Clarke2023,Deng2023,LlabresPohl2023} and marine ecosystems \citep{Jagerbrand2019, Marappan2022,Clovis2024}, and~increases safety \citep{Eliopoulou2023, AllianzCommercial2024} and security \citep{Bueger2024} risks along the dense shipping routes. Several mechanisms are in place to ensure traffic safety and to monitor compliance with local and international regulations in coastal waters and the open seas. Among~these are the Vessel Monitoring System, used in fishing, and~the Long-Range Identification and Tracking system, LRIT. Unlike these mechanisms, the~Automatic Identification System (AIS) allows public tracking of most operating vessels at high temporal resolution, including information such as vessel identity, port of call, and~current draft \citep{Shepperson2018}. The~International Convention for the Safety of Life at Sea (SOLAS) requires ships of 300 gross tonnage (GT) and above on international voyages (or cargo ships of more than 500 GT on domestic voyages) and all passenger ships to regularly report voyage status information via so-called Class A type AIS transceivers (AIS-A) in intervals of a few seconds to several minutes \citep{IMO1974}. Also, other vessels can voluntarily use the AIS via the weaker and less prioritized Class B system (AIS-B). The~transmitted data can then be picked up by any suitable radio receiver on land, other vessels (which can relay the signal to land), or~by satellites. Although~AIS was intended primarily for the immediate safekeeping of maritime traffic, collected AIS data have been used in numerous scientific studies of activity patterns at sea \citep{Natale2015,Kroodsma2018,Cerdeiro2020, Yan2020,Visky2024}.

Governmental and private actors with access to space-based systems and dense terrestrial receiver networks can track vessel activity through the AIS with comprehensive coverage, but~access to these data is mostly restricted. In~contrast, community-based networks and some government authorities increasingly provide open access to their real-time or historical data from terrestrial AIS receivers \citep{Tu2018}. Using open access data has recently caught the attention of the research community \citep{Hadjipieris2025,Chen2025}. However, data from open access services are subject to their own distinct limitations. The range of picking up AIS signals from land is usually limited to several tens of nautical miles, restricting terrestrial receiver networks to only track vessels in coastal waters \citep{Emmens2021}. Moreover, AIS networks may lack receiver stations in certain coastal areas or may be regionally limited. Lastly, community-based networks usually do not apply strict control over the quality or origin of their data, making these networks vulnerable to the injection of accidentally or intentionally incorrect data \citep{Androjna2023,Kessler2024,Louart2024}.

This work demonstrates that vessel activities can nevertheless be comprehensively studied with data from open-access AIS networks, with~accuracy competitive to previous studies based on proprietary data. Accordingly, describing a suitable methodology in this paper will facilitate monitoring, controlling, or~studying maritime activities for any public or private stakeholder. In~doing so, we present two novel aspects in the analysis of AIS positioning data: first, we provide a rigorous estimate of uncertainties arising from incomplete data. This is of particular importance when data is regionally limited and of inhomogeneous coverage. Second, we provide vessel counts and number densities for moving vessels and stationary activities, such as fishing or berthing in ports, throughout the work in consistent units. This allows us a direct comparison of both contributions to the overall maritime activity, and~to  infer port locations complementary to previous methods based on clustering~algorithms.

We demonstrate the feasibility of providing competitive insights from open-access AIS data using three months of vessel activity in the Baltic Sea in 2024, an~area with diverse shipping activity that is mostly, but~not exclusively, composed of coastal waters. Moreover, numerous studies based on AIS data have been published for this region \citep{Jalkanen2014,Loptien2014,Goerlandt2017,Lensu2019,FernandezArguedas2018,HELCOM2018,Kulkarni2022,Dettner2023}, facilitating the comparison with previous works. We emphasize that we do not address intentional AIS data manipulation, and~caution is required when using community data for safety or security purposes. This particularly applies to the Baltic Sea, where cases of illegitimate deactivation of AIS transceivers and AIS spoofing have been repeatedly reported~\citep{Caprile2024}.

The remainder of this paper is organized as follows. \Cref{sec:methods} presents our methods. \Cref{sec:data_acquisition} introduces the study area and describes the AIS data acquisition; \Cref{sec:data_filtering} outlines the cleansing applied to the AIS message data; and \Cref{sec:journey_reconstruction} outlines our model for converting the AIS data into a spatio-temporal representation of vessel journeys. \Cref{sec:vessel_metrics} describes the derivation of vessel metrics, activity maps, and~port locations. In~\Cref{sec:results_uncertainties}, we quantify the uncertainties in these metrics. \Cref{sec:results} presents the results of the Baltic Sea analysis. We discuss our findings and conclude in \Cref{sec:discussion}.


\section{Methods}
\label{sec:methods}
\unskip

\subsection{Analysis Region and Data~Acquisition}
\label{sec:data_acquisition}

The Baltic Sea is one of the busiest maritime areas in the world. Between~8\% \citep{Serry2014} and 15\% \citep{Hakkinen2012} of the worldwide cargo is estimated to pass through the region. Cargo shipping through the Baltic Sea increased by about 2\% yearly in the decade before 2014~\citep{Boteler2015} and between 2015 and 2019 \citep{Statista2021}. Overall maritime transport volume is expected to continue increasing until 2030 \citep{Matczak2018,Friman2019}, driven by both commercial and cruise shipping \citep{FernandezArguedas2018}. Surrounded by nine littoral states, excluding Norway, the~Baltic Sea also forms a maritime zone with large geopolitical~significance.

\begin{figure}[t]
\centering
	\includegraphics[width=0.52\textwidth, trim=0px 0px 0px 15px, clip]{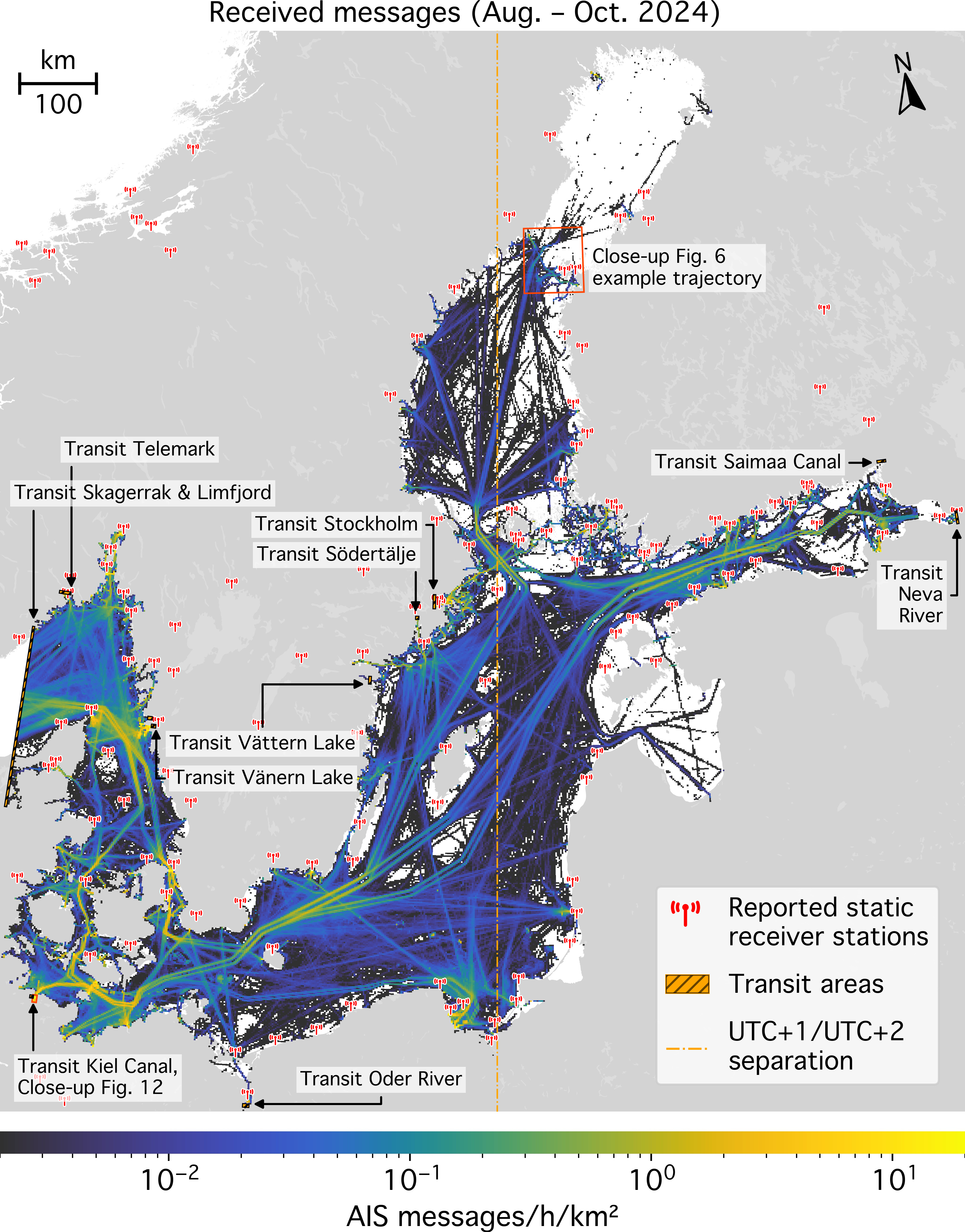}
	  \caption{\textls[15]{
Reported positions of the AIS-A messages received in the Baltic Sea ROI between 29~July and 27~October \textls[25]{2024. The~figure spans from $9^\circ\,\mathrm{E}$ to $32^\circ\,\mathrm{E}$ in longitude, and~from $53^\circ\,\mathrm{N}$ to $66^\circ\,\mathrm{N}$ in latitude. The~11 defined transit areas are also shown, along with~the static receiver stations reported by the network. The~orange dash-dotted line marks the longitude $\lambda=19.5^\circ\,\mathrm{E}$, used to assign messages to the UTC+1 (westwards) and UTC+2 (eastwards) time zones. All maps in this paper are shown in the Transverse Mercator projection.} }}
      \label{fig:roi_data_map_baltic}
\end{figure}

Different geological and political definitions exist for the extent of the Baltic Sea. According to the Baltic Marine Environment Commission (Helsinki Commission, HELCOM), the~Baltic Sea is the basin between Scandinavia and the European Plain, limited in the west by the latitude of $\varphi=57.741^\circ\,\mathrm{N}$ between Skagen and Gothenburg at the transition from the Kattegat to the Skagerrak \citep{HELCOM1992}. For~this analysis, we choose a larger area bounded in the west by the longitude $\lambda$~=~$9^\circ\,\mathrm{E}$ through the Skagerrak and Limfjord in Denmark, and~separated from inland waters by nine additional transit areas, as~defined in \Cref{tab:exit_areas} and shown in \Cref{fig:roi_data_map_baltic}. This region of interest (ROI) covers an area of $440{,}492\,\SI{}{\square\km}$ and two time zones, UTC+1 and UTC+2, which we divide at $\lambda=19.5^\circ\,\mathrm{E}$ (Figure~\ref{fig:roi_data_map_baltic}).

For the defined ROI, we subscribed to the open AIS data stream from \citet{aisstream2025} for 91 days between Monday, 29 July 2024, 00:00:00~UTC and Sunday, 27 October 2024, 23:59:59~UTC, receiving position-report messages from moving vessels (navigational status zero) and ship static data from all AIS-A vessels, collecting a total of $91{,}111{,}731$ messages from $14{,}620$ different Maritime Mobile Service Identity (MMSI) numbers. We discarded all messages corresponding to coordinates on land. We also do not include Class B vessels in our analysis because of their large seasonal variability, dominantly associated with leisure vessels, and~the poorer data quality. However, we separately collected messages from Class B vessels to estimate their average contribution to the overall shipping activity in the uncertainty discussion of \Cref{subsec:aisb}. We did not collect position reports of different navigational statuses and use the static reports to trace a vessel during a stationary~period.

\begin{figure}[t]
    \includegraphics[width=.95\textwidth, trim=0px 0px 0px 15px, clip]{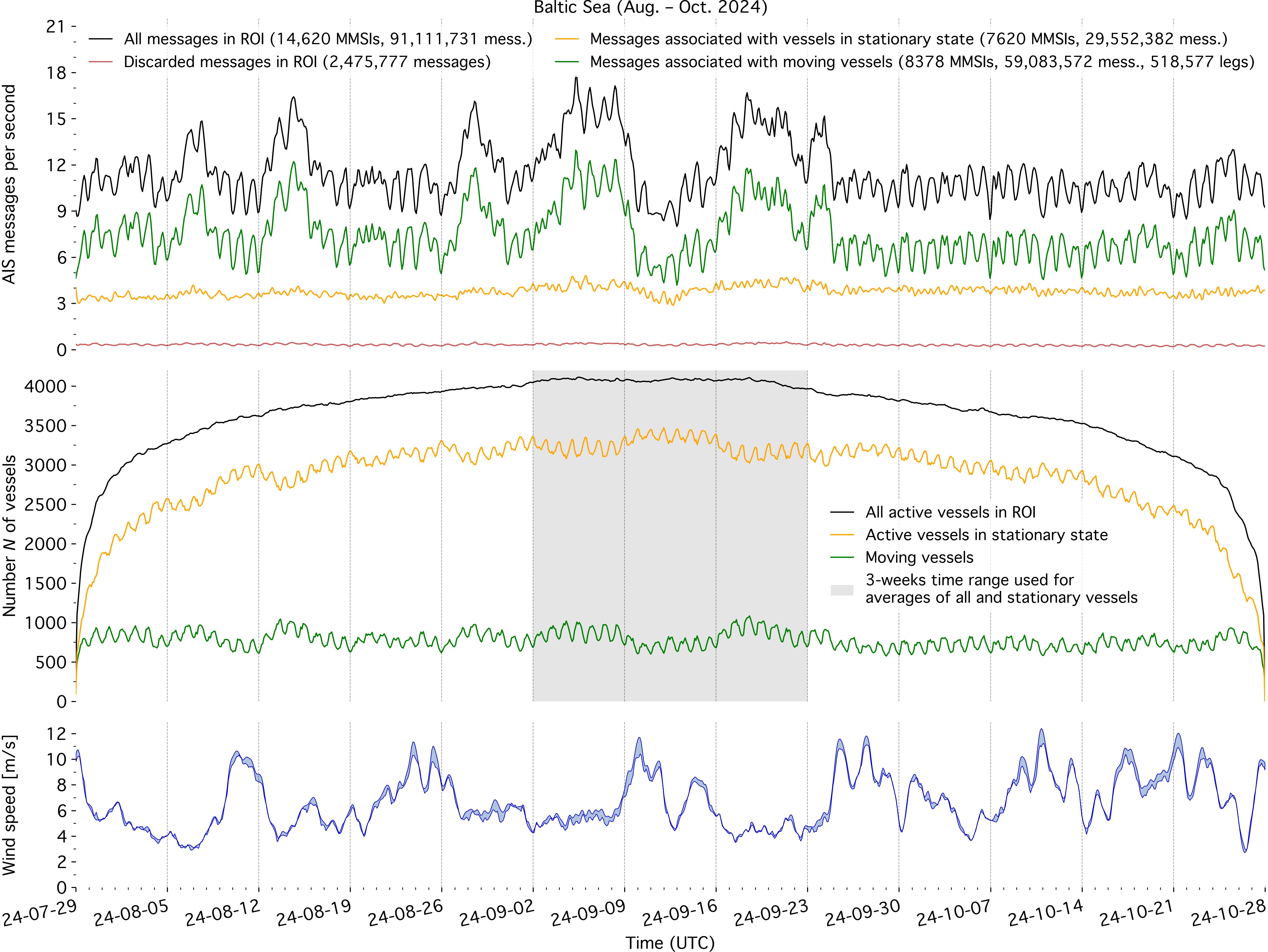} 
	  \caption{
Timeline of received AIS-A message counts (\textbf{top panel}), inferred number of vessels operating in the ROI at the given time (\textbf{middle panel}), and~prevalent wind conditions (\textbf{bottom panel}) during the analysis period. The~vertical dotted lines denote the beginning of each Monday. The~black curves in the top and middle panels represent the sum of the other curves. The~gray-shaded three-week interval in the middle panel is used to derive time-averaged estimates of the total number of vessels and stationary vessels in the ROI (see \Cref{sec:vessel_metrics,sec:results} for details). Wind conditions are shown as the range of the initial values given by the \citet{Copernicus2024} and \citet{GFS2024} forecast models averaged over the ROI area.
Figures~\ref{fig:hist_times_messages_cart_baltic} and \ref{fig:hist_times_vessels_cart_baltic} show the top and middle panels on a daily cycle.}

    \label{fig:ais_signals_time_baltic}
\end{figure}

We saved the messages with the reported timing information (one-second precision); latitude and longitude with decimeter precision (six decimal places); the MMSI number; the reported speed over ground with two decimal places (position reports only); and~vessel type and destination (static reports only), corresponding to a compressed data size of 1.01 GB (12~bytes per message). No gap larger than \SI{400}{\second} in the received message timestamps was experienced during the analysis period (\Cref{fig:ais_signals_time_baltic}, top panel). 

Figure~\ref{fig:roi_data_map_baltic} shows the spatial distribution of the vessel positions in the received messages. We also collected status report messages from 147 static base stations of the network in the surrounding area and show their locations in Figure~\ref{fig:roi_data_map_baltic}. It can be seen that, in particular, the Gulf of Riga and the Gulf of Bothnia (the northern Baltic Sea) are poorly covered by receiver stations, resulting in low message counts in these areas. Additionally, a~decrease in message density is observed between the island of Gotland and the Latvian coast, indicating poor reception by the nearest receivers of the messages sent from this area. In~Figure~\ref{fig:ais_signals_time_baltic}~(top panel), the~black curve shows the 
distribution of messages according to their reported times. From~the message rate associated with stationary vessels (yellow curve, see \Cref{subsec:movement_splitting} for details), we infer that we obtained a stable data stream from the subscribed~service. 

\begin{figure}[t]
  \centering
    \includegraphics[width=0.5\textwidth, trim=-18px 0px 0px 15px, clip]{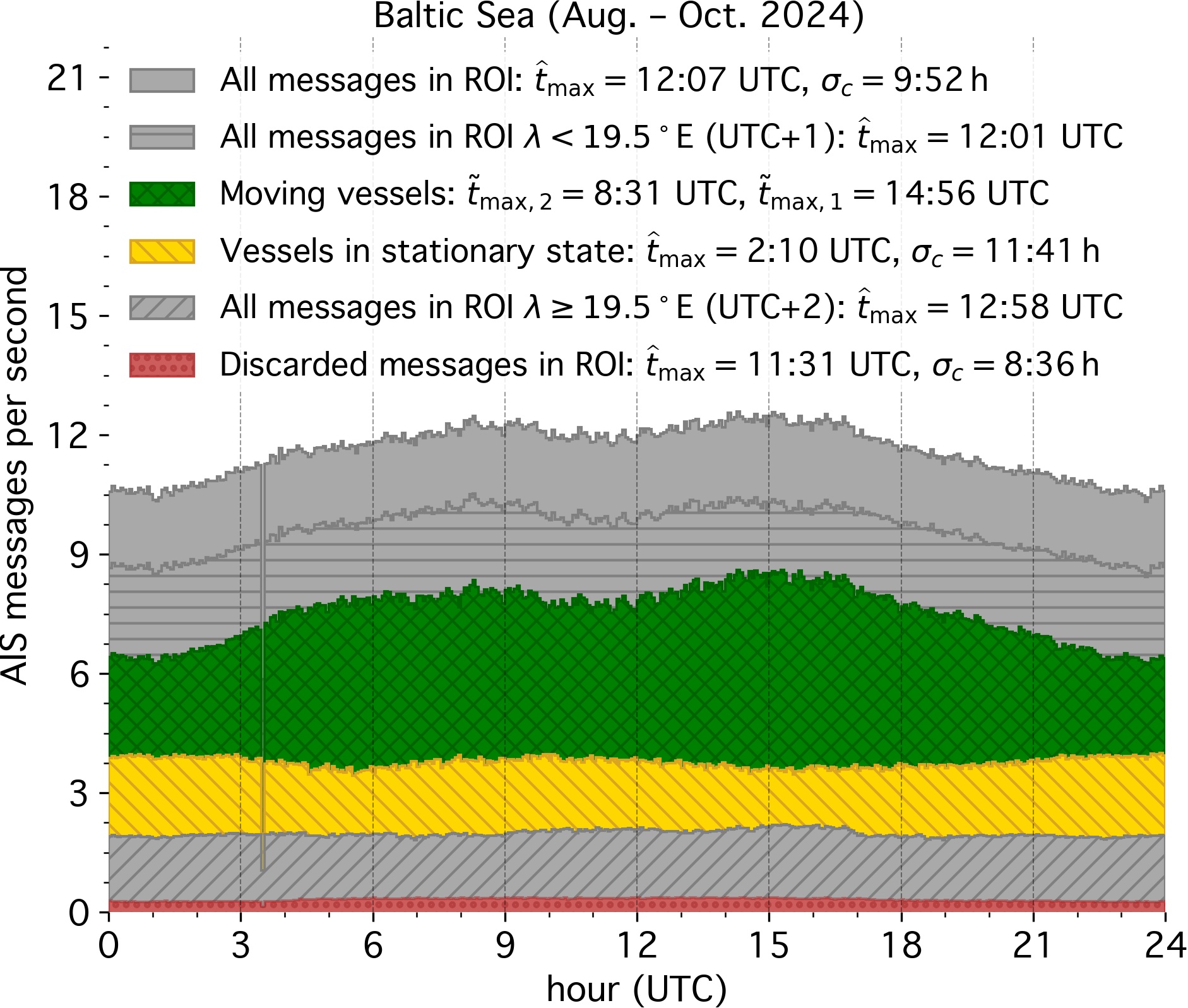} 
        \caption{
    AIS  message activity throughout the day, stacked over the 91-day analysis period (Figure~\ref{fig:ais_signals_time_baltic}, top) using a bin width of \SI{4}{\minute}. The times $\hat{t}_\text{max}$ and $\tilde{t}_\text{max}$ denote the modes and circular means, and~$\sigma_\text{c}$ the circular standard deviations \citep{Mardia1999}. The~sharp gap at 3:30 UTC is a data provider feature.
    } 
    \label{fig:hist_times_messages_cart_baltic}

    \vspace{20pt} 

    \includegraphics[width=.5\textwidth, trim=0 -1.75mm 0 15px, clip]{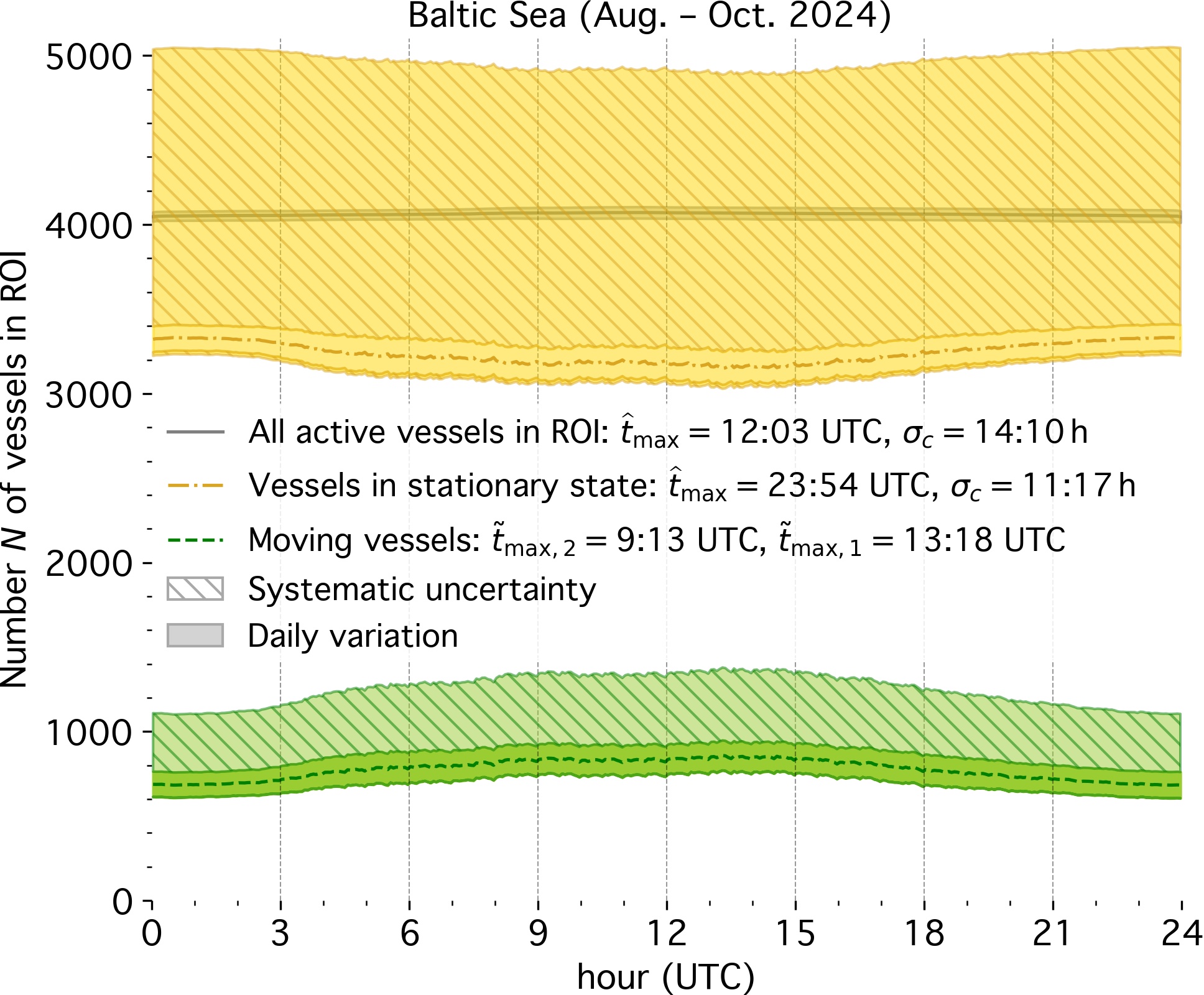} 
    \caption{
    Vessel activity throughout the day. Systematic uncertainties (hatched bands) include the analysis uncertainty and estimates for additional untracked and AIS-B vessels, and~are upwards dominated by AIS-B vessels. 
        } 
    \label{fig:hist_times_vessels_cart_baltic}

\end{figure}
\unskip


\subsection{AIS Message Cleansing and~Classification}
\label{sec:data_filtering}

Numerous studies have addressed AIS data cleansing \citep{Zhao2018,Wolsing2022}, with~recent works proposing both rule-based \citep{Lv2023,AsianDevelopmentBank2023} and machine-learning  \citep{Zhang2023a,Yang2024} approaches. In~this work, we reconstruct vessel movements from the four AIS message variables of reported latitude, longitude, timestamp, and~MMSI number to distinguish vessels. We then apply a rule-based algorithm~to: 
\begin{enumerate}
    \item Discard messages considered erroneous or irrelevant;
    \item  Classify messages as corresponding to a vessel being in motion, stationary, or~entering or leaving the ROI---we jointly refer to the latter two states as transits; 
    \item Separate the moving and stationary periods. 
\end{enumerate}

 We assume a unique MMSI for each vessel. However, we discuss a possible bias due to MMSI sharing during the uncertainty estimation (\Cref{subsec:mmsi_sharing}).

\subsubsection{Removal of Static Vessels from the~Analysis}
\label{subsec:static_removal}

We remove all vessels whose positions remain within a square of side length of \SI{400}{\meter} during the entire 91-day period. This scale is chosen as the analysis's spatial resolution, similar to the previous analysis by \citet{HELCOM2018} and matching the available bathymetry data from \citet{NOAA2022}. This removal defines the term ``active vessel'' in the remainder; we consider only vessels that move at least once. We choose this approach because, for~a vessel appearing static and present only within a limited time, it is impossible to determine whether it is moored at the given position throughout the analysis period or only during the time range of received messages. This removal also eliminates spurious messages with a single, most likely scrambled MMSI number. This cut removes only 1\% of all messages but 37\% of all observed MMSI numbers.

\subsubsection{Correction of Positions in the Ship Static~Data }
\label{subsec:position_correction}

An AIS-A vessel is required to issue a static voyage report every \SI{6}{\minute}, for~which the used data stream includes the position information of the last position report. We incorporate these static messages into the position reports; however, we correct the position coordinates under the assumption of constant speed and straight vessel movement between the enclosing position reports. This correction affects 18\% of all messages (i.e., the~percentage of static reports). After~this, we compute vessel speeds between all subsequent messages based on the distances and time~differences. 

\subsubsection{Removal of Duplicate Messages at Low~Speeds}
\label{subsec:duplicate_removal}

\begin{figure}[t] 

		\includegraphics[width=.98\textwidth, trim=0px 0px 0px 15px, clip]{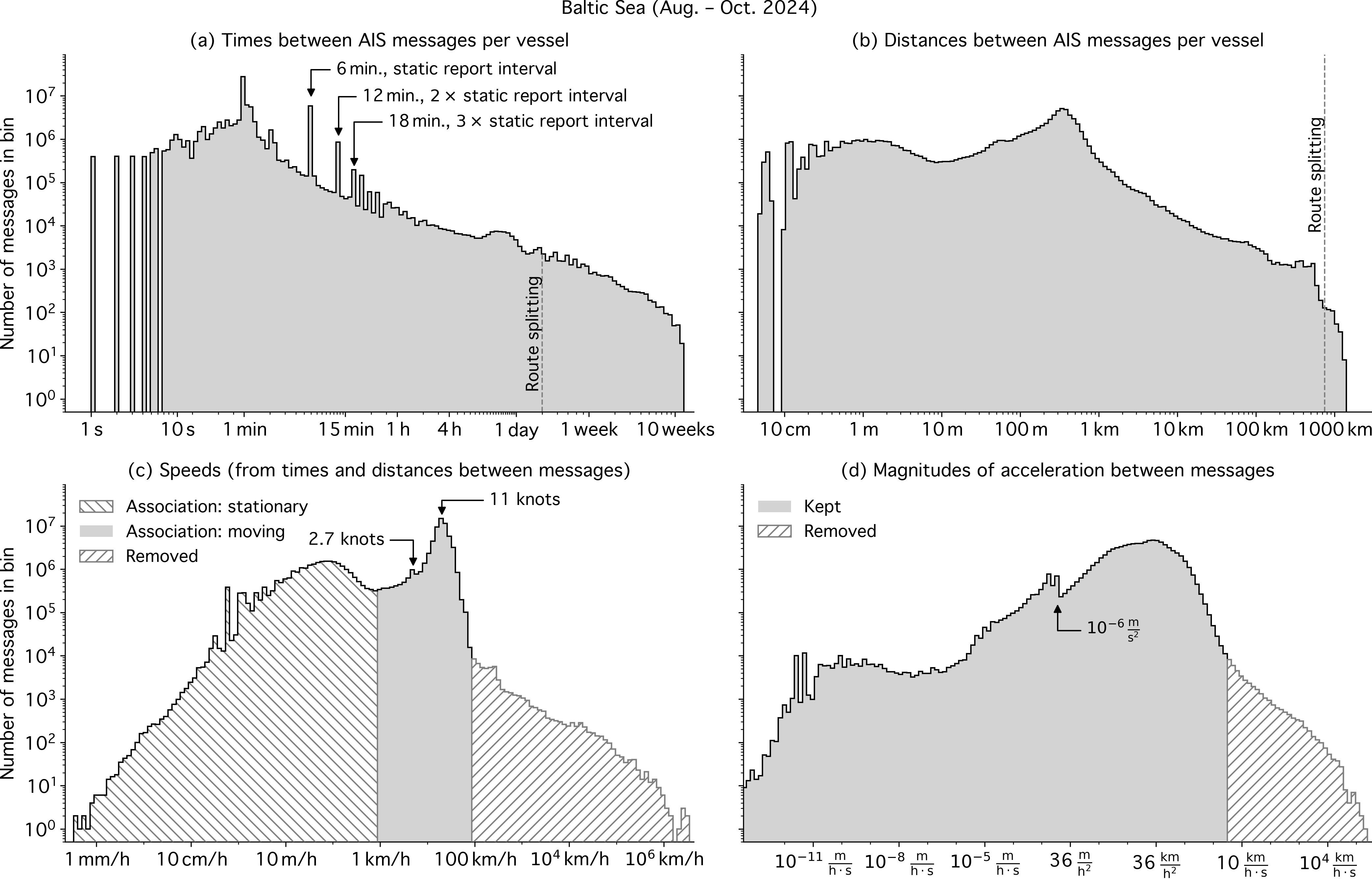} 
	  \caption{
Messages per vessel retained after the cleaning steps in Sections  \ref{subsec:static_removal}--\ref{subsec:duplicate_removal}, sorted into logarithmic intervals based on the time differences between subsequent messages (\textbf{a}), distances (\textbf{b}),  speeds (\textbf{c}), and~accelerations (\textbf{d}) between those messages. The~step in the acceleration histogram around $10^{-6}\,$\SI{}{\meter/\square\second} is due to the static-report interval multiples and decimeter-length precision.
    }
      \label{fig:ais_signals_histograms_baltic}
\end{figure}

To reduce the amount of data, we also remove messages that report, at high frequency, insignificant vessel movement. We remove messages that occur within \SI{5}{\second} of a previous message, while the speed is below \SI{1}{\km/\hour} (the threshold for a stationary-vessel state motivated below) and the distance between the messages is less than \SI{1}{\m}. This removes an additional $0.7\%$ of the messages. 
Messages retained after this step are shown in \Cref{fig:ais_signals_histograms_baltic}. Figure~\ref{fig:ais_signals_histograms_baltic}a shows that most messages are broadcast in intervals of \SI{1}{\minute}. The~static-report interval is also visible, as~are its `overtones', indicating occasional missed message broadcast or reception. Figure~\ref{fig:ais_signals_histograms_baltic}b shows that the most prominent time interval corresponds to a dominant message distance of $\sim$\SI{350}{\meter}, or~correspondingly, a~dominant speed of \SI{11}{\knot} (Figure~\ref{fig:ais_signals_histograms_baltic}c). 
Figure~\ref{fig:ais_signals_histograms_baltic}d shows the messages grouped by relative acceleration. Note that  three~neighboring messages are always used to assign an acceleration value to the central~one.

\subsubsection{Movement~Segmentation}
\label{subsec:movement_splitting}
Various methods have been proposed to identify the state of motion of a vessel and to separate moving periods from vessel stops \citep{AsianDevelopmentBank2023,VanDerWielen2024,Wijaya2024,Iphar2024,Guo2024,Qiang2025}. 
In this work, we adopt a rule-based, iterative splitting algorithm to sort messages into periods of vessel movement and into stationary or absent phases. Stationary phases correspond to vessels being moored at a constant position or slowly drifting. Absent phases correspond to vessels that have left, but~re-enter the ROI.
 We apply the splitting algorithm twice: once before removing messages based on speed and acceleration constraints, and~again after. We do this to disentangle, as~far as possible, the~segmentation from the cleansing. We split between those messages where at least one of the following conditions is~met:
\begin{itemize}
    \item The speed is lower than \SI{0.5}{\knot} ($\sim$\SI{1}{\km/\hour}). This value is commonly acknowledged in the literature as a threshold to discriminate stationary from moving vessels \citep{HELCOM2018,Cerdeiro2020,Li2022}. Also, it is motivated by our data, where the distribution in Figure~\ref{fig:ais_signals_histograms_baltic}c shows a prominent minimum in log-space at $\sim$\SI{0.5}{\knot}.
    \item  The time gap is larger than \SI{48}{\hour} (displayed by the dashed vertical line in Figure$\,$\ref{fig:ais_signals_histograms_baltic}a), an~estimate of the longest duration to travel a straight path through the Baltic Sea.
    \item The distance is greater than \SI{750}{\km} (displayed by the dashed vertical line in Figure~\ref{fig:ais_signals_histograms_baltic}b), also an upper bound for traversing a clear line of sight through the ROI.
    \item The vessel track passes through a transit area. These transit areas, defined in \autoref{tab:exit_areas},  are located at the ROI boundary and are shown in Figure~\ref{fig:roi_data_map_baltic}.
\end{itemize}

If the splitting affects subsequent messages in a row, and~only due to the speed condition, those messages are classified as belonging to a stationary period of a vessel. Single messages separated by a time or distance split (i.e., outliers in time or space) are removed. If~a vessel track passes through a transit area, messages are split at the largest time difference between two subsequent messages within that area, including the last message before entering and the first after leaving the area. If~a movement, enclosed by idle times before and after, entirely falls into a transit area, the~messages are~removed. 

This first pass classifies 31\% of the  messages in the ROI to belong to stationary vessel states (approximately corresponding to the left hatched area in Figure~\ref{fig:ais_signals_histograms_baltic}c),
removes an additional 0.3\% of the messages,
and creates 898{,}222 separate~movements.

\subsubsection{Removing Outliers by Implausible Speeds and~Accelerations}
\label{subsec:cut_speed_acceleration}
Among the messages classified to belong to vessel movements, we remove erroneous data indicating physically implausible speeds and accelerations between the messages. We remove the second message of any two between which the speed exceeds \SI{50}{\knot} (Figure~\ref{fig:ais_signals_histograms_baltic}c). This value is above the maximum cruising speed at which current high-speed vessels operate under normal conditions \citep{Yun2012}. Also, we remove any message with an associated absolute acceleration larger than \SI{1}{\meter/\square\second} (Figure~\ref{fig:ais_signals_histograms_baltic}d), acknowledging that under normal conditions, vessels rarely experience horizontal accelerations larger than $0.1$ g,  the~gravitational acceleration. Because~the speed and acceleration between the remaining messages change after each removal, cleaning is performed iteratively until no more outliers are detected. The~cleaning removes an additional 0.4\% of the messages,
indicated by the right hatched areas in Figure~\ref{fig:ais_signals_histograms_baltic}c,d. It also removes all but the first of multiple vessels sharing an MMSI. We discard all messages removed by this cleaning for the main analysis. However, we repeat the full analysis pipeline for the events removed in this step to assess the impact of possible physically separate vessels sharing the same MMSI (see Section~\ref{subsec:mmsi_sharing}).

\subsubsection{Second Segmentation Pass and Area~Filtering}
\label{subsec:second_pass}
After cleaning from infeasible speeds and accelerations, we apply the splitting algorithm a second time, adding  dozens of new movements,
thousands of stationary messages,
and removing seven messages. We then apply the area cut from \Cref{subsec:static_removal} to each individual movement, reducing again the number of movements by 311{,}665 short-distance tracks. However, this time, we do not remove the corresponding $\sim$$10^6$ messages, but~classify them as belonging to stationary vessel periods.

\subsubsection{Combining~Movements}
\label{subsec:combining_movements}

Finally, we iteratively combine all movement periods that are separated by $\leq$$\SI{120}{\second}$, twice the most prevalent reporting interval, and~remove the first message of the attached movement. This is to undo false segmentation caused by messages whose position coordinates were not updated since the previous reporting time, hence suggesting a short rest of the vessel. It also recombines uninterrupted movements through a transit area that remain within the ROI. This combination is performed in 68{,}002 cases (12\% of the remaining movements), reducing the number of movements and increasing the number of discarded messages by this amount. 
This finally leaves us with 65\% of all messages in the ROI associated with 518{,}577 movements,
32\% of the  messages associated with stationary periods in between,
and 3\% discarded messages.
During the various cleansing steps, 6242 MMSI numbers were removed from the original count, leaving us with 8378 vessels for which to reconstruct their~journeys.

The top panel of Figure~\ref{fig:ais_signals_time_baltic} shows the messages associated with vessel movements (green curve) or stationary periods (yellow curve), as~well as the removed messages (red curve). It can be seen that the irregularly fluctuating message rate is almost entirely associated with movement periods, while the rate associated with stationary vessels remains largely stable throughout the analysis period. \Cref{fig:hist_times_messages_cart_baltic} shows the same data on a daily~cycle.


\subsection{Vessel Journey~Model}
\label{sec:journey_reconstruction}
After cleansing and classifying the messages, we build logical entities that allow us to infer the vessel position and state at any time and to compute vessel metrics and spatially resolved activity maps. As~part of this process, we simplify the message information to a given precision in position and time. For~the remainder of the paper, we use the following terminology: We define a route as the geometric track of a vessel's movement between two~endpoints without timing information. A trajectory refers to the movement of a vessel along a route, including timing and speed information. Subsequent trajectories of a particular vessel, separated by stationary periods or times absent from the ROI, form the legs of its journey. Each vessel undertakes exactly one journey within the analysis time~range. 

\subsubsection{Route~Simplification}
\label{subsec:route_reconstruction}

We infer the vessel routes from the message positions during the movement periods, reducing the position data to a subset of tracklets, connected at waypoints, where the vessels change course. Dedicated implementations and extensions of the Ramer--Douglas--Peucker (RDP) simplification algorithm \citep{Ramer1972,Douglas1973} exist for this task \citep{Pallero2015,Tang2021a,Mehri2025}.
Also, new algorithms have been published for AIS data \citep{Onyango2022,Zhang2024}. However, we find that sufficient data compression while maintaining accuracy at any required precision is achieved using the fast implementation of the standard RDP algorithm by \citet{HugelSimplification2021}. We express the algorithm's simplification threshold $\epsilon$ to determine the required route precision in terms of the maximum tolerance distance $d_{\text{tol}}$ from any original message position as 
\begin{align}
    \epsilon = \frac{2\times d_{\text{tol}}}{60\,\mathrm{nmi}}\times \cos \overline{\varphi}\,,
    \label{eq:rdp_threshold}
\end{align}
with $\overline{\varphi}$ the mean latitude coordinate of the route waypoints. Equation (\ref{eq:rdp_threshold}) converts metric distance to angular distance (in units of degrees) and applies a less stringent $\epsilon$ for routes at higher latitudes, where larger longitude ranges represent the same metric distance. The~additional factor of 2 assumes that large portions of routes were originally straight paths and ensures that they are reconstructed when positions fluctuate around both sides of the path. In~the Baltic Sea at $\cos(60^\circ) = 0.5$, both scaling factors mostly cancel out. The~performance of this threshold conversion is shown in \cref{fig:model_analysis_baltic} (middle panel). For this analysis, we set $d_{\text{tol}}=\SI{100}{\meter}$, a~quarter of the defined spatial analysis~resolution.

\subsubsection{Speed~Model}
\label{subsec:speed_model}
To express the number densities of moving vessels, it is necessary to know the vessel speed at any moment in time. Therefore, along the simplified routes, we use speed control points describing the vessel motion. We define these points independently of the route waypoints, at~irregular spacing, only at significant speed changes. We consider a significant speed change as a 5\% variation in vessel speed, estimated at a central message position as the average over the interval between the preceding and following messages. Between~speed control points, we interpolate speeds linearly in time, corresponding to constant acceleration. Also, we normalize all speed points such that the full trajectory duration matches the time difference between the first and last message in the vessel~movement.

The simplified route waypoints, together with the speed model, form a trajectory model. We compared the vessel positions and times in the model with the original AIS data and found that the model, despite its simplicity, describes the vessel motion over time with an accuracy of about 1\%. The~detailed comparison of the simplified trajectory model with the AIS data points is provided in 
Appendix~\ref{app:speed_model_accuracy}.

\Cref{fig:trajectory_model_example,fig:speed_model_example} illustrate this modeling for a single trajectory. In~this example, the~route never deviates more than \SI{100}{\meter} from the original messages (teal diamond dots in Figure~\ref{fig:speed_model_example}, bottom left), and~when additionally considering timing information, model positions differ at most by \SI{800}{\meter} or \SI{80}{\second} from the messages (black circles in the bottom panels of Figure~\ref{fig:speed_model_example}). In~Figure~\ref{fig:speed_model_example} (top panels), speeds calculated from positions and timestamps (green diamonds) show a larger fluctuation than the values reported with the AIS messages (red circles). This is due to the one-second timestamp precision and position inaccuracies. Nevertheless, the~speed model (blue squares) is successfully inferred from the noisy~data.

\begin{figure}[t!]
\centering
		\includegraphics[width=0.5\textwidth]{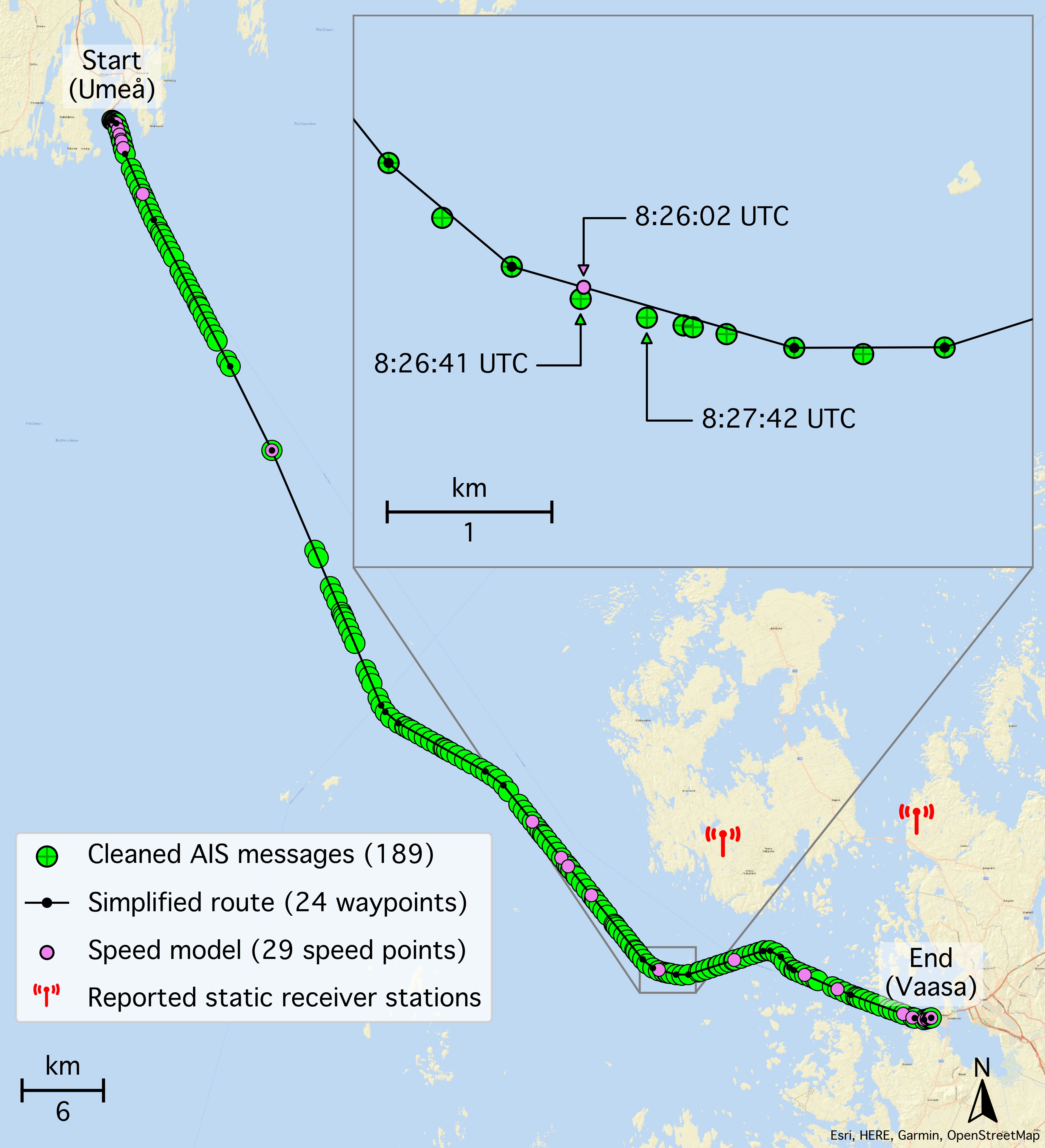}
	  \caption{An  example of the reduction in original AIS messages (large green dots) into a simplified trajectory model. The~model is defined by route waypoints (small black dots) and speed control points (violet dots) along the route. This example depicts a ferry trip from Ume\aa{} to Vaasa, departing on 2024/07/29 at 7:52:24 UTC+2 and arriving at 12:22:00 UTC+3. The~AIS messages and speed points correspond to those in Figure~\ref{fig:speed_model_example}, parametrized by time and distance.
      }
      \label{fig:trajectory_model_example}
    \vspace{30pt}  
		\includegraphics[width=.98\textwidth]{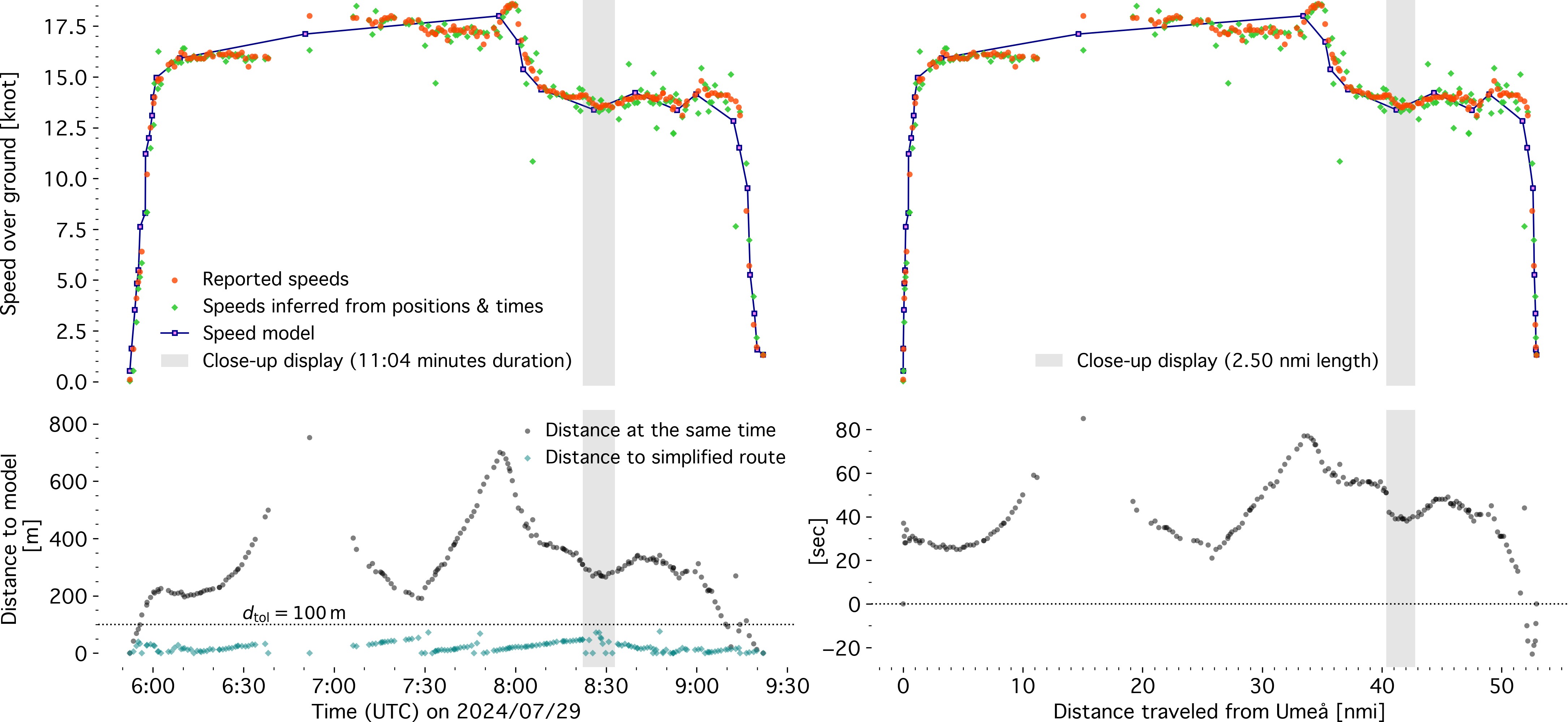} 
	  \caption{
Ferry speed during the example trajectory shown in Figure~\ref{fig:trajectory_model_example} over time (\textbf{top left}) and distance (\textbf{top right}) from the departure port. Both panels compare reported speeds from AIS data (red circles), speeds inferred from the reported positions and timestamps of the AIS messages (green diamonds), and~our speed model (blue squares with connecting curves). (\textbf{bottom left}) Distances of message positions to the route track (teal diamonds) and to the vessel on the model trajectory at the same time (black circles). (\textbf{bottom right}) Time differences between message timestamps and the times when the ferry is at the closest position on the simplified route. Gray-shaded areas indicate the zoomed-in section in Figure~\ref{fig:trajectory_model_example}.
    }
      \label{fig:speed_model_example}
\end{figure}

\subsubsection{Journey~Construction}
\label{subsec:journey_construction}
Finally, we combine the trajectories into logical units of vessel journeys, including stationary or absent times between the vessel movements. 
The challenge of this step is to determine, for~each interval between two movements, whether a vessel remains stationary at the position connecting the movements, or~whether it is temporarily absent from the ROI. This is because vessels may anchor close to the ROI boundary rather than leave, especially in areas near ports. Conversely, due to imperfect receiver coverage, vessels may leave and re-enter the ROI without messages being detected within a transit area at the ROI boundary. Also, although~vessels are expected to transmit position and static reports even when not moving, we observed that this is not always the case. Sometimes, vessels switch off their transceivers while moored, even if not supposed to do so \citep{Emmens2021}. 
To correctly classify stationary or absent periods, we apply the following conditions if a movement starts or ends in a transit~area:
\begin{enumerate}
    \item[1a.] If more than three messages are received from a vessel between two movements, we assume it stays within the ROI at all times between the movements.
    \item[1b.] In the transit areas except the Skagerrak and Kiel Canal, we only consider a vessel as temporarily leaving the ROI if the time between two movement periods exceeds
\begin{align}
    t_\text{thr} = t_0\times \left(\mathrm{max}(v_\text{exit}, v_\text{entry})\,/\,10\,\mathrm{kn}\right)^{-4}\,,
    \label{eq:t_thresh}
    \end{align}    
    where $v_\text{exit}$, $v_\text{entry}$ are the last (first) observed speed of the previous (following) leg.  
    Using $t_0=\SI{6}{\hour}$, we flag vessels at speeds $> \SI{44}{kn}$ as temporarily absent for message gaps $< \SI{1}{\min}$, the~most common reporting interval (Figure~\ref{fig:ais_signals_histograms_baltic}a). 
    The exponent in Equation~(\ref{eq:t_thresh}) reduces the threshold by more than one order of magnitude for doubled speed and is chosen such that high-speed vessels ($\gtrsim \SI{40}{kn}$) are always flagged as transiting.
    \item[2a.] In the Kiel Canal transit area and within \SI{400}{m} in front of it (one cell in the later chosen grid), we mark all vessels with idle time $>\SI{24}{\hour}$ as absent, overruling condition 1a.
    \item[2b.] In the whole ROI, we only consider a vessel present after receiving the first message in our analysis period, and~consider a vessel absent after the last message.
\end{enumerate}

Conditions 1a and 1b eliminate false-positive transit events. Conditions 2a and 2b reduce false-negative transit events, and~in particular, remove ``forgotten'' vessels in front of transit areas. Condition 2b results in an increase in counted stationary vessels at the beginning and a decrease at the end of the analysis period, 
as seen in the middle panel of Figure~\ref{fig:ais_signals_time_baltic}.
For all periods not flagged as absences, we define the idle position as the central point between the end of the previous and the start of the following vessel~movement.


\subsection{Computation of Maritime Activity~Metrics}
\label{sec:vessel_metrics}
\unskip

\subsubsection{Vessel Count and Transit~Rates}
\label{subsec:vessel_numbers_fluxes}

From the complete journey model for each of the 8378 vessels, we compute the number of vessels simultaneously present in the ROI, and~the rate of entering or leaving vessels per time. To~this aim, we divide the analysis period into 32{,}760 intervals of \SI{4}{\minute} each, and~for each interval, we count the number of moving or stationary vessels in the ROI, and~the first (or last) appearances of moving vessels flagged as absent before (or after). These counts are performed separately for six categories based on the vessels' self-reported type and two categories of vessel size. \Cref{tab:vessel_types} provides the mapping of all AIS vessel codes to these categories. We retrieved the vessel sizes in terms of gross tonnage (GT) from  \citet{VesselFinder2025} for 99\% of all observed vessels with an International Maritime Organization identification number (``IMO vessels''). However, IMO vessels only represent 65\% of all vessels in the~analysis.

The middle panel of Figure~\ref{fig:ais_signals_time_baltic} shows the count of simultaneously operating vessels in the ROI. The~edge effects for the count of stationary vessels (yellow curve) result from the fact that vessels entering from outside operate in the ROI for several weeks.  These effects span almost the entire analysis period;~therefore, we base our estimate of the average number of stationary vessels (and, by~extension, the~sum of moving and stationary vessels) on the central 21 days of the analysis~period. 

\subsubsection{Spatially Resolved~Maps}
\label{subsec:maps}
Based on the vessel journey models, we can express the overall vessel activity in terms of spatially resolved number density. Therefore, we pixelate the ROI into a regular grid with spacing $(\Delta\varphi,\,\Delta\lambda)=(15'',\,30'')$ in latitude~$\varphi$ and longitude~$\lambda$, our analysis resolution. In latitude direction, this results in a cell height $h$ of about \SI{464}{\meter}, and~in longitude, into~a width $w$ between \SI{559}{\meter} at $\varphi=53^\circ$ and \SI{378}{\meter} at $\varphi=66^\circ$, similar to the resolution from \citet{HELCOM2018}. On~this grid, for~the vessel movements, each time a vessel passes through a grid cell, we perform the following:
\begin{enumerate}
\item Count the occurrence of the cell crossing;
\item Compute the average speed during the crossing;
\item Compute the average bearing during the crossing; and~\item Compute the average duration of the crossing.
\end{enumerate}

In order to do so, we sample all vessel trajectories along geodesics between the waypoints in intervals of \SI{100}{\meter}, a~quarter of the cell width. If~a tracklet between two waypoints crosses land at an elevation higher than \SI{2}{\meter} above mean sea level according to \citet{NOAA2022}, we exclude the tracklet from the sampling. We choose a threshold of \SI{2}{\meter} to avoid excluding too many tracklets in the cleft coastal areas of the Baltic Sea. If~several consecutive samplings fall into the same grid cell, we assume the vessel to be constantly present within the cell, and~we assign a single crossing, the~mean speed, and~circular mean bearing for this crossing to the cell. For~a computationally feasible estimate of the crossing duration, we compute the crossing time from the average path length and the average vessel speed within the cell. The~average path length through a rectangular cell under a constant bearing angle $\alpha$ is defined by Equation (\ref{eq:line_segment_const_angle}). After~sampling an entire trajectory, we normalize the sum of crossing durations to the trajectory duration (with the duration of the excluded tracklets subtracted). Finally, after~having sampled all trajectories, we sum up all crossing occurrences and durations in each~cell. 

For the stationary periods, we compute the mooring durations of each vessel in the corresponding cells. However, we do not spatially resolve stationary periods where the distance between the previous leg end and the following leg start coordinates exceeds \SI{1}{\km}, due to the uncertainty of the actual vessel position during the idle~period. 

We then sum up the movement times and stationary durations and divide by the analysis period duration (91 days) and cell area to obtain the average number of vessels per unit~area.

\subsubsection{Finding Port~Areas}
\label{subsec:ports}

The combined stationary and moving vessel density allows us to estimate port areas complementary to previous works. Various techniques have been described to identify anchoring and port locations,~using spatial clustering algorithms \citep{AsianDevelopmentBank2023,Liu2023,Iphar2024,Qiang2025,Hadjipieris2025}, a~kernel density approach \citep{Millefiori2016}, or~machine learning on training data \citep{Yan2022}. In~this work, we infer port areas from the pixelized number density through a two-step map segmentation. First, we find seed locations as the barycenter positions of areas enclosed by a threshold of $0.5\,$vessels$/\SI{}{\square\km}$, with~Gaussian smoothing over $1.5\,$cells ($\SI{\sim\,700}{m}$) to denoise the density. Using these seed locations, we determine port areas with a watershed algorithm \citep{scikit-image}, applied to water areas above half the threshold value with the same smoothing, and~the port locations as barycenter positions within these areas. The~density threshold is the single pre-determined parameter of the algorithm. For~the purpose of this study, it is chosen such that we obtain a median area size of about \SI{4}{\square\km} for a single port. Additionally, we dilate all areas to a size of at least three cells ($\SI{\sim\,0.7}{\square\km}$). From these areas, we also estimate vessel arrivals by counting all journey legs that end within port areas (but do not start within the same areas). For 
 the Kiel port area, containing the Kiel Canal transit area, we exclude vessels that enter the canal, but~vessels can arrive in the fjord from the canal.


\subsection{Uncertainties in the Vessel~Metrics}
\label{sec:results_uncertainties}
\unskip

\subsubsection{Uncertainty About the Receiver~Coverage}
\label{subsec:uncert_coverage}

Quantifying incomplete and inhomogeneous AIS coverage is generally a challenge for AIS data analysis \citep{Natale2015}. In~principle, reconstructed trajectories should not depend on message frequency along vessel routes. However, this does not hold when trajectories cannot be constructed at all due to insufficient message data, or~when critical waypoints marking course direction changes are missing. Within~the ROI, poor data coverage leads to an underestimation of the relative share of moving vessel time in the total vessel count. While we lack comparison data to quantify this uncertainty, we assume the impact on the total vessel count to be minor. We attempt to quantify the impact of poor coverage at the ROI boundaries on vessel counts and transit rates, as described~below. 

\subsubsection{Uncertainty About Leaving or Moored~Vessels}
\label{subsec:uncert_leaving}

If a vessel is falsely flagged as moored within the ROI while it is in reality leaving and reentering (or vice~versa), it confounds both the count of stationary vessels and the vessel transits. The movement may have been detected as interrupted by the time cut in \Cref{subsec:movement_splitting}, but~is falsely identified as an idle period instead of an~absence.

In \Cref{subsec:journey_construction}, we applied rules to reduce such false-positive and false-negative transit counts. To~estimate the uncertainty of false-positive and false-negative transits, we create two variations in our default set (case \textit{df}) of vessel journeys: a variation further reducing false positives (case \textit{hi}), and~a variation reducing false negatives, but~allowing for more false positives (case \textit{low}).
\begin{itemize}
    \item For the Skagerrak and Kiel Canal, we vary the transit areas in which to check for absence in the journey construction of \Cref{subsec:journey_construction}, using a region smaller (case \textit{hi}) and larger (case \textit{low}) than the default. A~larger area reduces false-negative transits, but~may increase false positives (see \Cref{tab:exit_areas} for all area definitions).
    \item For all other transit areas, we only create a looser condition on false positives by setting $t_0=\SI{1}{\hour}$ in Equation (\ref{eq:t_thresh}) (case~\textit{low}).
\end{itemize}

Case \textit{low} results in lower stationary-vessel numbers and higher transit rates, and vice~versa for case \textit{hi} (detailed in \Cref{tab:case_comparison}). We include this variation in our overall uncertainty estimate individually for each 
vessel 
category, with uncertainties up to 17\% (cargo vessels) for the vessel counts, and~up to 55\% (Kiel Canal) for the all-vessel transit rates. Due
 to low statistics, the~transit-rate uncertainty  becomes 100\% for some categories. 

\subsubsection{Uncertainty About MMSI~Sharing}
\label{subsec:mmsi_sharing}
Multiple simultaneously operating vessels may share the same MMSI (e.g.,~\citep{Pallotta2013,Zhao2018}). In~our analysis, all of those except the one with the first detected message would be discarded in the cleansing. To~assess the possibility of having removed valid vessels with duplicate MMSI numbers, we repeated the full analysis pipeline using only the messages removed in the cleansing step, Section \ref{subsec:cut_speed_acceleration}. 
After scrutinizing the obtained vessel journeys, we found that all checked trajectories correspond to the same as in the retained data, and~that they are built from messages rejected due to time jitters, etc.

For an estimate of the largest possible error, we obtain from the analysis of rejected messages a total travel time of 7104$\,$\SI{}{\day} on valid tracklets versus 60{,}577$\,$\SI{}{\day} in the main analysis. With~this, we would underestimate vessel activity by at most 10\% 
due to rejecting vessels with duplicate MMSI. If~we assumed the same message frequency for all vessels, this value would reduce to at most the 3\% rejected messages. In~summary, we assume that the error due to MMSI sharing is negligible, consistent with the analysis from the supplementary material in \citet{Kroodsma2018}, and~do not include it in our combined uncertainty~estimate.

\subsubsection{Uncertainty About Vessels Not Using the~AIS}

\begin{figure}[H]
    
    \subfloat[\centering\label{fig:ratio_of_all_tracked_vessels_baltic}]{\includegraphics[width=0.23\textwidth]{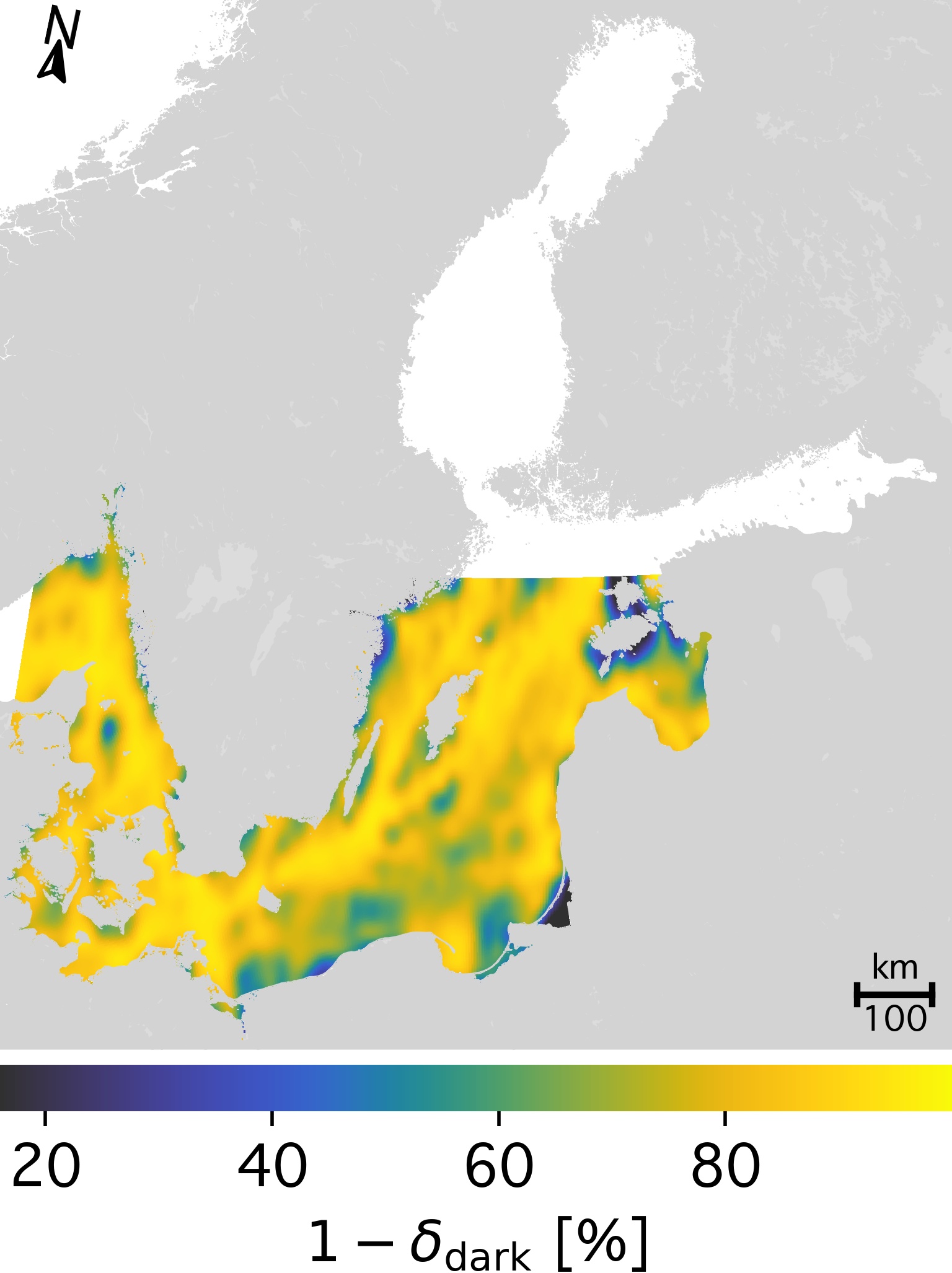}}
    \hfill
    \subfloat[\centering\label{fig:ratio_of_fishing_vessels_among_all_vessels_baltic}
]{\includegraphics[width=0.23\textwidth, trim=-0.25mm 0 0 0, clip]{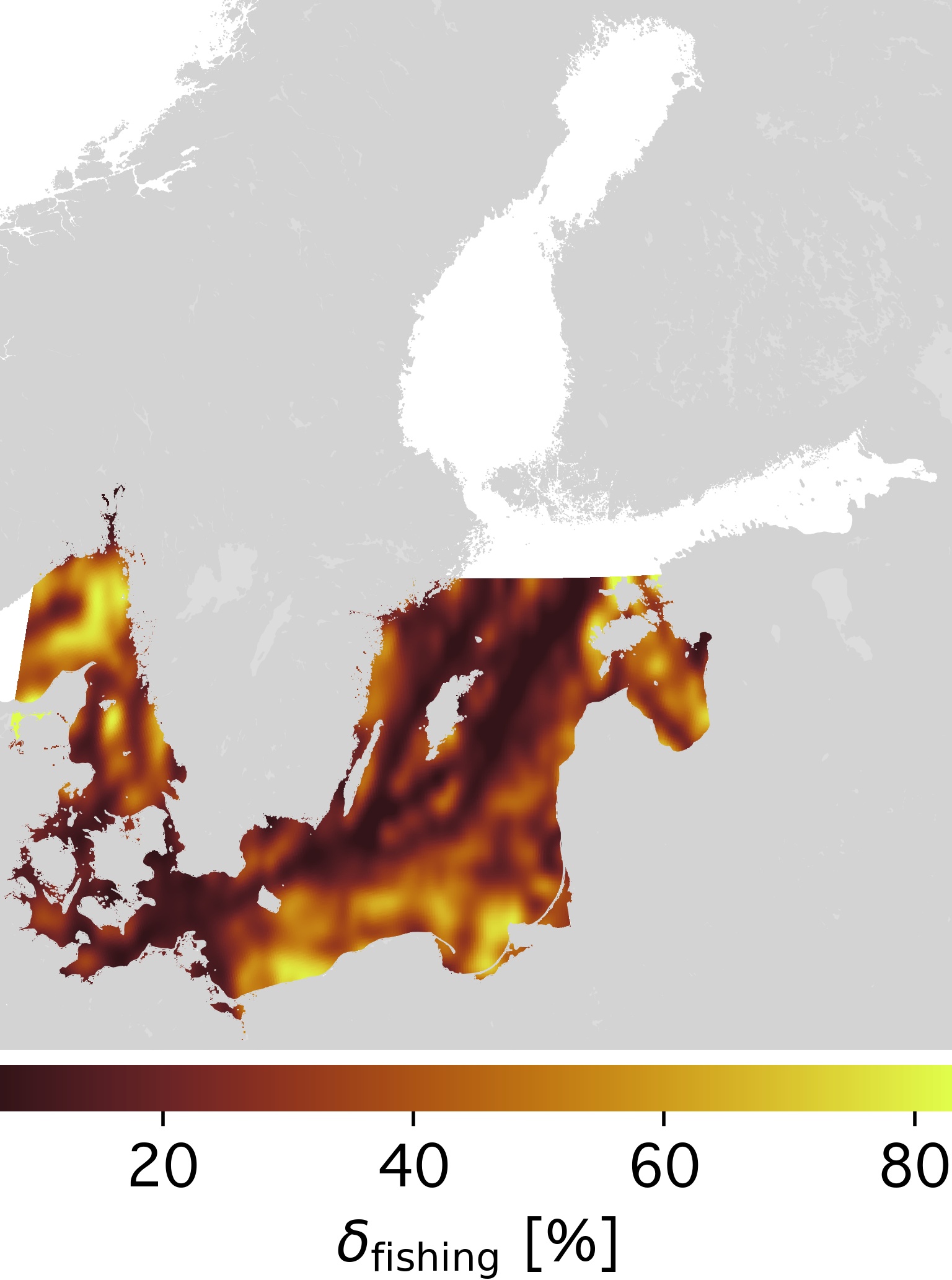}}
\hfill
    \subfloat[\centering\label{fig:ratio_of_tracked_non_fishing_vessels_baltic}
]{\includegraphics[width=0.23\textwidth]{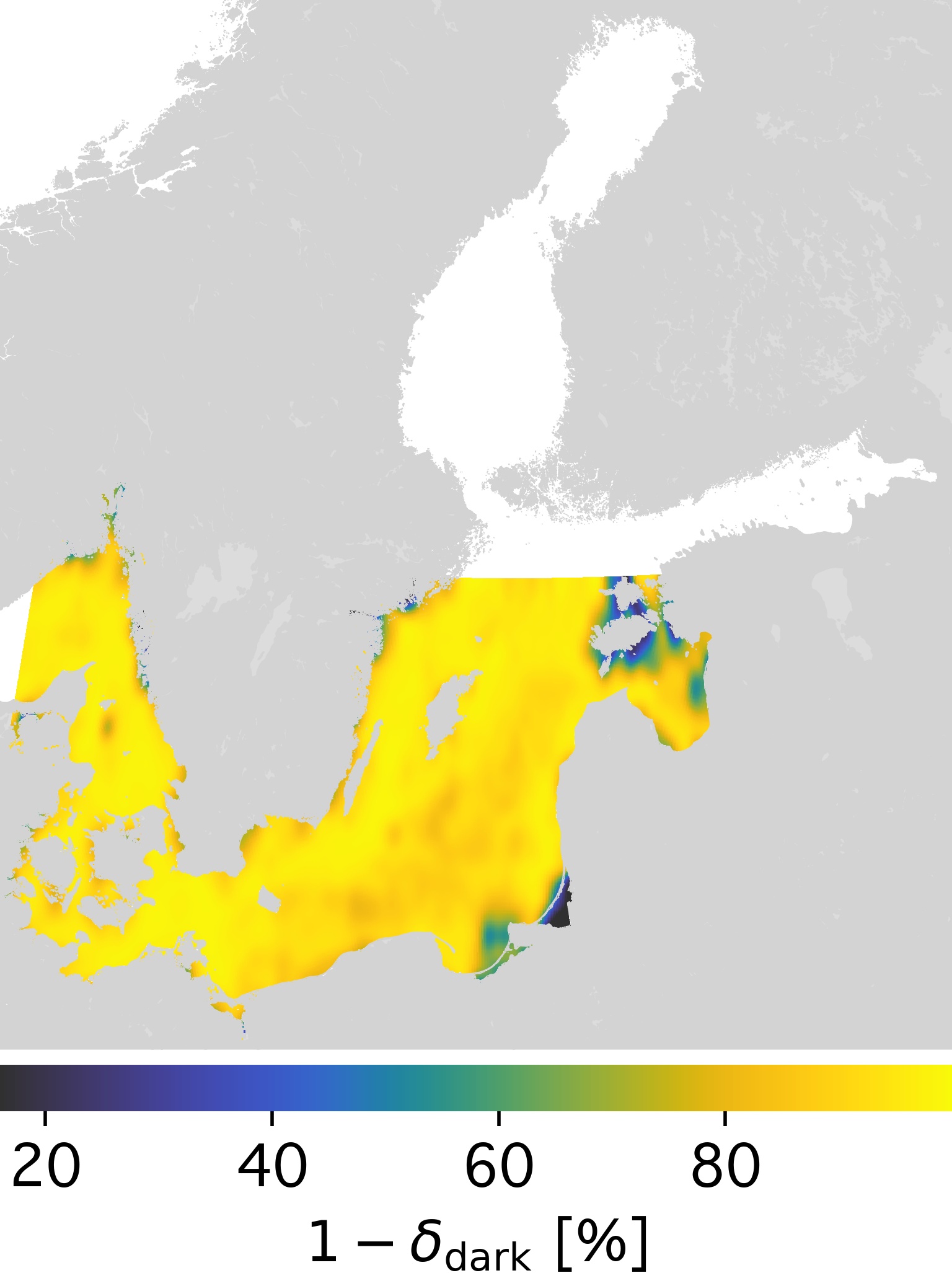}}
    \hfill
    \subfloat[\centering\label{fig:ratio_of_tracked_fishing_vessels_baltic}
]{\includegraphics[width=0.23\textwidth]{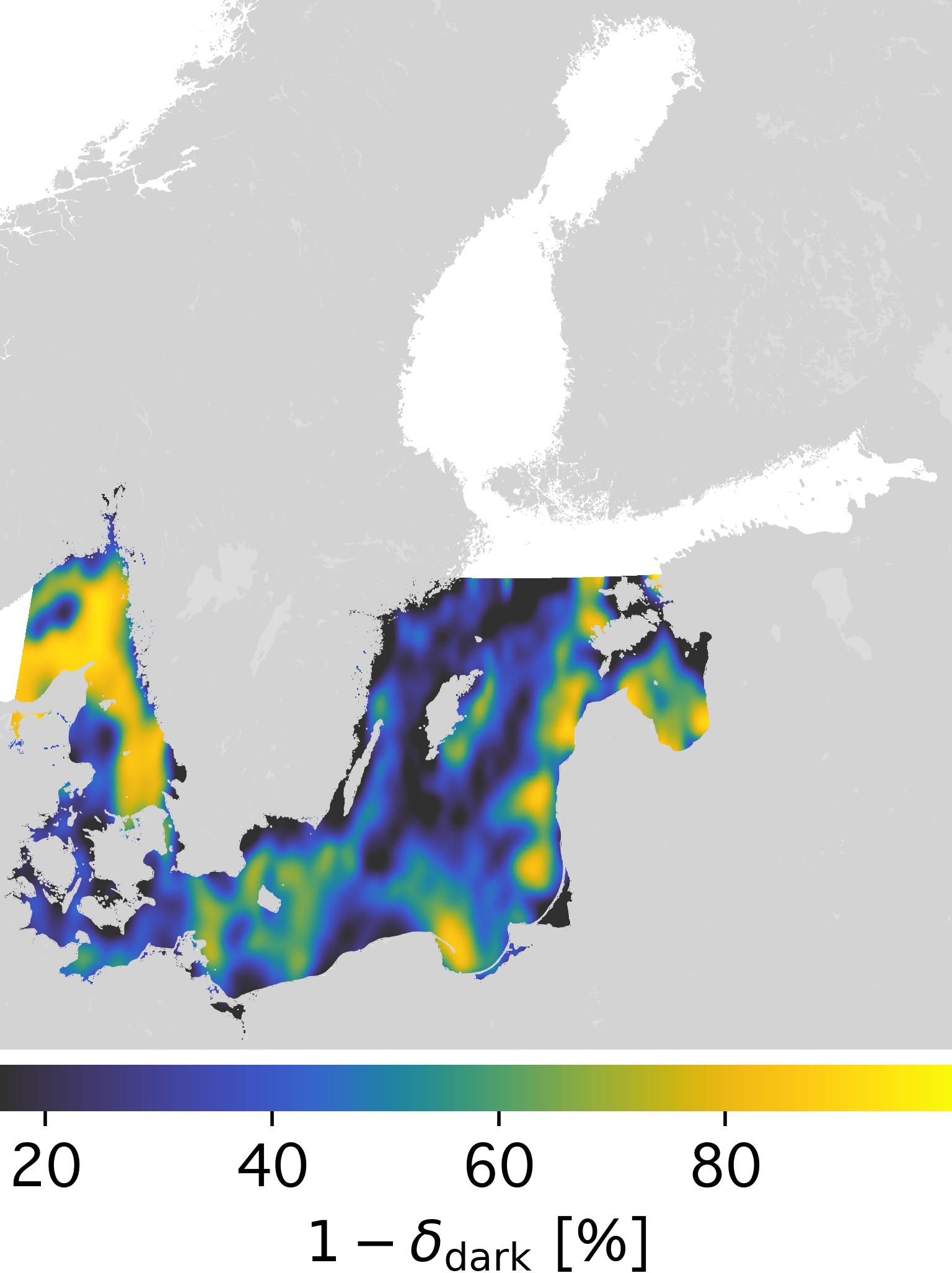}}
    \caption{Fraction of vessels tracked by the AIS and the fraction of fishing vessels in the Baltic Sea between 2017 and 2021, based on data from \citet{Paolo2024}. Paolo~et~al. only analyze the region $\varphi\leq59.1^\circ\,\mathrm{N}$ within the Baltic~Sea. (\textbf{a}) Fraction of tracked vessels among all vessels. (\textbf{b}) The fraction of fishing vessels among all vessels. (\textbf{c}) The fraction of tracked vessels among non-fishing vessels. (\textbf{d})~The fraction of tracked vessels among fishing vessels.}
    \label{fig:tracked_vessels}
\end{figure}

Some vessels do not submit any AIS signals, either because they are not required, miss having their transceiver switched on, or~intentionally, legitimately or illegitimately, prefer to operate unnoticed \citep{Emmens2021}. \citet{Paolo2024} conducted a comprehensive correlation of worldwide satellite imagery with reported AIS positions in coastal waters. In~\Cref{fig:tracked_vessels}, we reproduce their public data at $(\Delta\varphi,\,\Delta\lambda)=(0.2^\circ,\,0.2^\circ)$ resolution. From~their data (Figure~\ref{fig:tracked_vessels}a), we read
\begin{align}
\delta_\text{dark} =
\begin{cases} 
    &21\% \quad\text{(full ROI)}\\    &13\% \quad\text{(Skagerrak at $9^\circ\,\mathrm{E}$) }\\
    &\;\,0\% \quad\text{(Kiel Canal at Holtenau) }
    \label{eq:fraction_dark}
\end{cases}
\end{align}
the fraction of untracked vessels in the Baltic Sea and the Skagerrak Strait. \citet{Paolo2024} do not provide data in the direct vicinity of shorelines, including the Kiel Fjord, and~we assume that all vessels passing the Kiel Canal correctly use the AIS. \mbox{While \citet{Paolo2024}} find that the usage of the AIS is particularly poor among fishing vessels (Figure~\ref{fig:tracked_vessels}d), we include a global constant fraction, defined by Equation (\ref{eq:fraction_dark}), in~our combined uncertainty estimate for~simplicity.

\subsubsection{Uncertainty About AIS-B~Vessels}
\label{subsec:aisb}

As stated in \Cref{sec:data_acquisition}, our analysis excludes vessels using the AIS-B. We find from our data that in the summer months, about twice as many AIS-B vessels ($\gtrsim$40\% of all vessels) are operating in the Baltic Sea as in the winter months ($\sim$20\%), primarily attributed to leisure activity. On~a yearly average, we assume that a fraction of 30\% of all vessels using the AIS use Class B transceivers. For simplicity, we assume the same fractions for moving and stationary~vessels.

 We find that the fraction of all vessels that enter or leave the Baltic Sea and use the AIS-B is lower, at $21\%$, than~their share among the activities within the ROI. The~percentage of AIS-B vessels is slightly larger (22\%) among the traffic through the Skagerrak than through the Kiel Canal (20\%). The percentages for each vessel category used in the uncertainty estimates for vessel numbers and transit rates are provided in \Cref{tab:vessel_aisb}.

\subsubsection{Combined Uncertainty of Vessel~Counts}

We include in our uncertainty estimates the bracketing of false-positive/negative transits and the underestimation of vessels due to missing untracked and AIS-B vessels. We build  upper ($+$) and lower ($-$) bounds $\Delta N_\text{syst}$ and $ \Delta \dot{N}_\text{syst}$ on vessel numbers $N$ and transit rates $\dot{N}$ as
\begin{equation}
    \begin{aligned}
    \left(\Delta N_\text{syst}^+\right)^2\!&= \left(N_\text{df} - N_\text{hi}\right)^2 + \left(\tilde{\delta}^2_\text{dark}\! + \tilde{\delta}^2_\text{ais-b}\right) N_\text{df}^2\,,\\
    \Delta N_\text{syst}^-\quad\!&= \;\, N_\text{df} - N_\text{low}\,, \\
    \left(\Delta \dot{N}_\text{syst}^{+}\right)^2\!&= \left(\dot{N}_\text{df}-\!\dot{N}_\text{low}\right)^2\!+ \left(\tilde{\delta}_\text{dark}^2\!+ \tilde{\delta}^2_\text{ais-b}\right)\!\dot{N}_\text{df}^2\,,\\
   \Delta \dot{N}_\text{syst}^{-} \quad\!&= \;\,\dot{N}_\text{df} - \dot{N}_\text{hi} \,,
    \end{aligned}
    \label{eq:syst_uncertainty}
\end{equation}
with $N$, $\dot{N}$, and~$\delta_\text{ais-b}$ according to each category under scrutiny, and $\delta_\text{dark}=21\%$ except for the transit rates $\dot{N}$ through the Skagerrak and Kiel Canal according to Equation (\ref{eq:fraction_dark}), and~$\tilde{\delta} = {\delta}/{(1-\delta)}$. In~all tables and figures, systematic uncertainties are provided according to Equation (\ref{eq:syst_uncertainty}).


\section{Results}
\label{sec:results}
\unskip

\subsection{Vessel Count and Density Within the Baltic~Sea}
\label{subsec:results_vessel_numbers}

We find an average total of 4061 vessels operating in the Baltic Sea at any moment in time between August and October 2024, divided into 774 moving vessels and 3287~stationary vessels (\Cref{tab:vessel_numbers}). Stationary vessels include ships with negligible speed over ground, such as those moored or drifting while engaged in fishing activities. Compared to previous estimates, this number is about double the figure reported by \citet{Storgard2012}, who claim ``nearly 2000 ships operating at any given moment [in the Baltic Sea]'' and is 7\% higher compared to the more recent estimate by \citet{Czermanski2017}, albeit in our case, on~a larger ROI extending into the Skagerrak. For~fishing vessels alone, \citet{Kroodsma2018} find an average of 729 vessels in our analysis ROI between 2012 and 2016, 
compatible with our count of 1028 vessels in the category ``Others including fishing'' (Table~\ref{tab:vessel_numbers}). Restricting fishing activity to self-reported AIS code 30 (codes 1002 and 1003 were not observed) results in an average of 378 active fishing vessels in our analysis. 
Figure~\ref{fig:ais_signals_time_baltic} (middle panel) shows these numbers over the 91-day analysis period. The~total number of vessels (black curve) remains largely constant over both the course of a day and week, with~the relative contribution of moving (green curve) and stationary vessels (yellow curve) strongly fluctuating during the day. Additionally, the~movement state correlates with the average wind speed in the Baltic Sea (Figure~\ref{fig:ais_signals_time_baltic}, bottom panel). For~moving vessels (green curve in Figure~\ref{fig:ais_signals_time_baltic}, middle panel), we do not observe any significant long-term variation on our 91-day~timescale. 

\Cref{fig:hist_times_vessels_cart_baltic} maps the same vessel count shown in Figure~\ref{fig:ais_signals_time_baltic} onto a daily scale. The~average total vessel count changes by only 0.4\% throughout the day, peaking around noon UTC and being slightly lower at night (gray solid line). However, the~proportion of moving vessels (green dashed line) varies by more than 20\%, with~modal peak activity occurring shortly after 1~p.m. UTC (2~p.m. in Central European Summer Time, 3~p.m. Eastern European Summer Time). A~shallower secondary peak appears in the morning around 9~a.m. UTC. We find that most vessel activity occurs in the western Baltic (accounting for 83\% of all vessel activity in terms of messages), as~does the fluctuation in daytime activity (Figure~\ref{fig:hist_times_messages_cart_baltic}). The~message rate from stationary vessels (yellow upper diagonally hatched distribution in Figure~\ref{fig:hist_times_messages_cart_baltic}) peaks at $\hat{t}_{\text{max}} = \text{2:10 UTC}$. Similarly, the~number of stationary vessels (dashed yellow curve in Figure~\ref{fig:hist_times_vessels_cart_baltic}) is largest at $\hat{t}_{\text{max}} = \text{23:54 UTC}$, 6\%~larger than at daytime. The~count of stationary and moving vessels only slightly varies between different days ($\sim$12\% or $\sim$90 vessels; Figure~\ref{fig:hist_times_vessels_cart_baltic}, yellow and green solid bands).
Uncertainties considered in \Cref{sec:results_uncertainties} result in a wide range of possible vessel counts. This combined uncertainty range is represented by the hatched bands in Figure~\ref{fig:hist_times_vessels_cart_baltic}. 
Therefore, vessel counts depend strongly on whether AIS-B vessels or even ships lacking AIS tracking devices are included in the~definition.

\Cref{tab:vessel_numbers,tab:vessel_numbers_gt} provide vessel counts categorized by type and their relative contributions. We find that cargo and tanker vessels (Table~\ref{tab:vessel_numbers}) spend most of their time moving, as~do large vessels (Table~\ref{tab:vessel_numbers_gt}). Conversely, vessel types such as pilot or tugboats, rescue vessels, or~fishing vessels are mostly found moored or~drifting.

\begin{table}[t] 
\caption{Average  number $N$ of vessels operating in the Baltic Sea (August--October 2024). The~first error terms represent statistical fluctuations based on day and time. The~second error terms account for systematic uncertainty regarding the ROI transits (both all vessels and stationary vessels) and the estimates on AIS-B and untracked vessels (positive terms only). The~table continues in Table~\ref{tab:vessel_numbers_gt}.}
\label{tab:vessel_numbers}
\scriptsize
\begin{adjustwidth}{-1.5cm}{0cm}
\newcolumntype{C}{>{\centering\arraybackslash}X}
\begin{tabularx}{\fulllength}{Ccccccc}
\toprule
  \textbf{\mbox{Vessel Type}}        &   \textbf{\mbox{All Active Vessels}} & 
  \textbf{\mbox{Moving Vessels}}     & 
  \textbf{\mbox{Stationary~Vessels}} & 
  \textbf{Percentage~(All)}          &
  \textbf{Percentage~(Moving)}       &
  \textbf{Percentage~(Stationary)}   \\
\midrule

All vessels in ROI & 
 $4061 \pm  4\,{}^{+   1639}_{-\;\,\;\;   22}$ & 
 $ 774 \pm 57\,{}^{+\;\,389}_{-\;\,\;\,\;\;0}$ & 
 $3287 \pm 62\,{}^{+   1639}_{-\;\,\;\;   22}$ & 
 $100\%$ & 
 $19\%$ &
 $81\%$ \\
\midrule

Passenger, high-speed & 
 $\;\,642 \pm  2\,{}^{+\;\,145}_{-\;\,\;\,\;\;1}$ & 
 $    137 \pm 38\,{}^{+ \;\,40}_{-    \;\,\;\,0}$ & 
 $\;\,505 \pm 37\,{}^{+\;\,145}_{-\;\,\;\;\;\,1}$ & 
 $16\%$ &
 $18\%$ &
 $15\%$ \\

Law enforcement, military & 
 $\;\,211 \pm    0.3^{   +\,71}_{      -      \,\;\,0}$ & 
 $\;\, 15 \pm \;\,5\,{}^{+ \;\,\;\,5}_{-    \;\,\;\,0}$ & 
 $\;\,196 \pm \;\,5\,{}^{+\;\,\;\,71}_{-\;\,\;\,\;\,0}$ & 
 $\;\,5\%$ &
 $\;\,2\%$ &
 $\;\,6\%$ \\

Cargo &
 $\;\,939 \pm     1\,{}^{+\;\,230}_{-\;\,\;\,10}$ & 
 $    338 \pm \;\,9\,{}^{+    102}_{- \;\,\;\,0}$ & 
 $\;\,601 \pm \;\,8\,{}^{+\;\,230}_{-\;\,\;\,10}$ & 
 $23\%$ &
 $44\%$ &
 $18\%$ \\

Pilot, tug, rescue, diving/dredging &
 $\;\,837 \pm  1\,{}^{+\;\,267}_{-\;\,\;\,\;\,5}$ & 
 $\;\, 54 \pm 10\,{}^{+ \;\,20}_{-    \;\,\;\,0}$ & 
 $\;\,783 \pm 11\,{}^{+\;\,267}_{-\;\,\;\,\;\,5}$ & 
 $21\%$ &
 $\;\,7\%$ &
 $24\%$ \\

Tanker &
 $\;\,404 \pm     1\,{}^{+\;\,\;\,82}_{-\;\,\;\,\;\,4}$ & 
 $    126 \pm \;\,1\,{}^{+    \;\,33}_{-    \;\,\;\,0}$ & 
 $\;\,278 \pm \;\,1\,{}^{+\;\,\;\,82}_{-\;\,\;\,\;\,4}$ & 
 $10\%$ &
 $16\%$ &
 $\;\,9\%$ \\

Others including fishing &
 $   1028 \pm  1\,{}^{+1152}_{-\;\,\;\,\;\;3}$ & 
 $    104 \pm 14\,{}^{+ 129}_{-    \;\,\;\,0}$ & 
 $\;\,924 \pm 18\,{}^{+1152}_{-\;\,\;\,\;\,3}$ & 
 $25\%$ &
 $13\%$ &
 $28\%$ \\
\bottomrule
\end{tabularx}
\end{adjustwidth}
\end{table}
\unskip

\begin{table}[t]
\footnotesize
\caption{Average 
 number $N$ of IMO vessels in the Baltic Sea and further categorized by gross tonnage (GT, August--October 2024). GT data are available for $99\%$ of observed IMO~vessels.}
\label{tab:vessel_numbers_gt}
\newcolumntype{C}{>{\centering\arraybackslash}X}
\begin{tabularx}{\textwidth}{lCcC}
\toprule
  & 
  \textbf{All IMO Vessels}                &
  \textbf{GT $\boldsymbol{<10{,}000}$}    &
  \textbf{GT $\boldsymbol{\geq10{,}000}$} \\ 
\midrule

All active vessels  & 
 $2265 \pm \;\,3\,{}^{+816}_{-\;\,19}$ & 
 $1672 \pm \;\,2\,{}^{+811}_{-\;\,17}$ & 
 $ 571 \pm \;\,1\,{}^{+ 86}_{- \;\,3}$ \\[0.2cm]
 
Moving vessels& 
 $\;\;648 \pm 22\,{}^{+323}_{-\;\,\;\,0}$ & 
 $\;\;399 \pm 30\,{}^{+252}_{-\;\,\;\,0}$ & 
 $    243 \pm 11\,{}^{+ 64}_{-    \;\,0}$ \\[0.2cm]
 
Stationary vessels & 
 $1587\pm 23\,{}^{+816}_{-\;\,19}$ & 
 $1273\pm 31\,{}^{+811}_{-\;\,17}$ & 
 $328 \pm 11\,{}^{+ 86}_{- \;\,3}$ \\
 \midrule
 
Percentage (all active vessels) &
 $56\%$       &
 $\geq$$41\%$ &
 $\geq$$14\%$ \\[0.15cm]
 
Percentage (moving vessels) &
 $84\%$       &
 $\geq$$52\%$ &
 $\geq$$31\%$ \\
 
Percentage (stationary vessels) &
 $48\%$       &
 $\geq$$39\%$ &
 $\geq$$10\%$ \\
\bottomrule
\end{tabularx}
\end{table}

\begin{figure}[t] 
    
    \subfloat[\centering\label{fig:density_map_baltic_all}]{\includegraphics[width=0.485\textwidth, trim=0px 0px 0px 15px, clip]{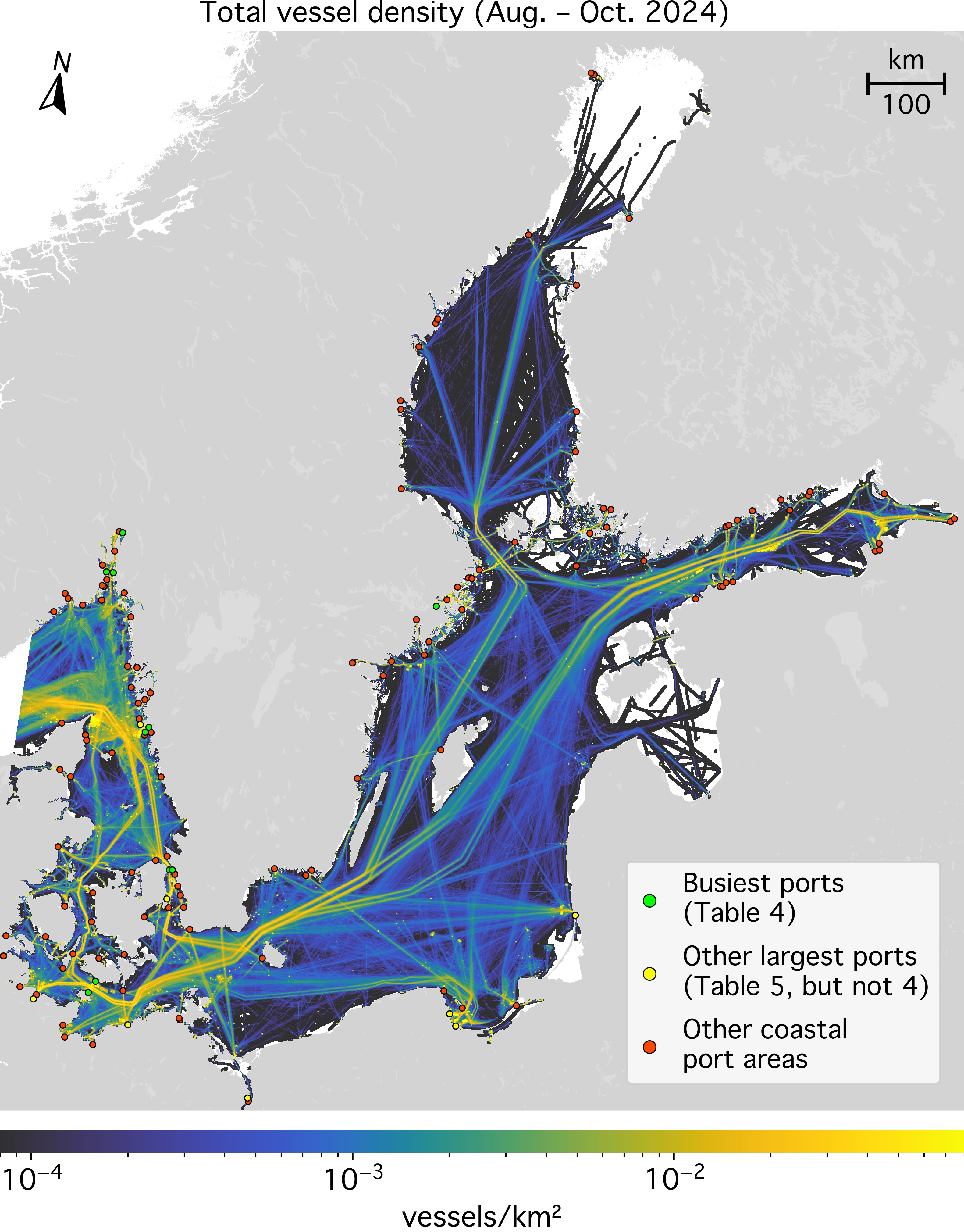}}
    \hfill
    \subfloat[\centering\label{fig:density_map_baltic_static}]{\includegraphics[width=0.485\textwidth, trim=0px 0px 0px 15px, clip]{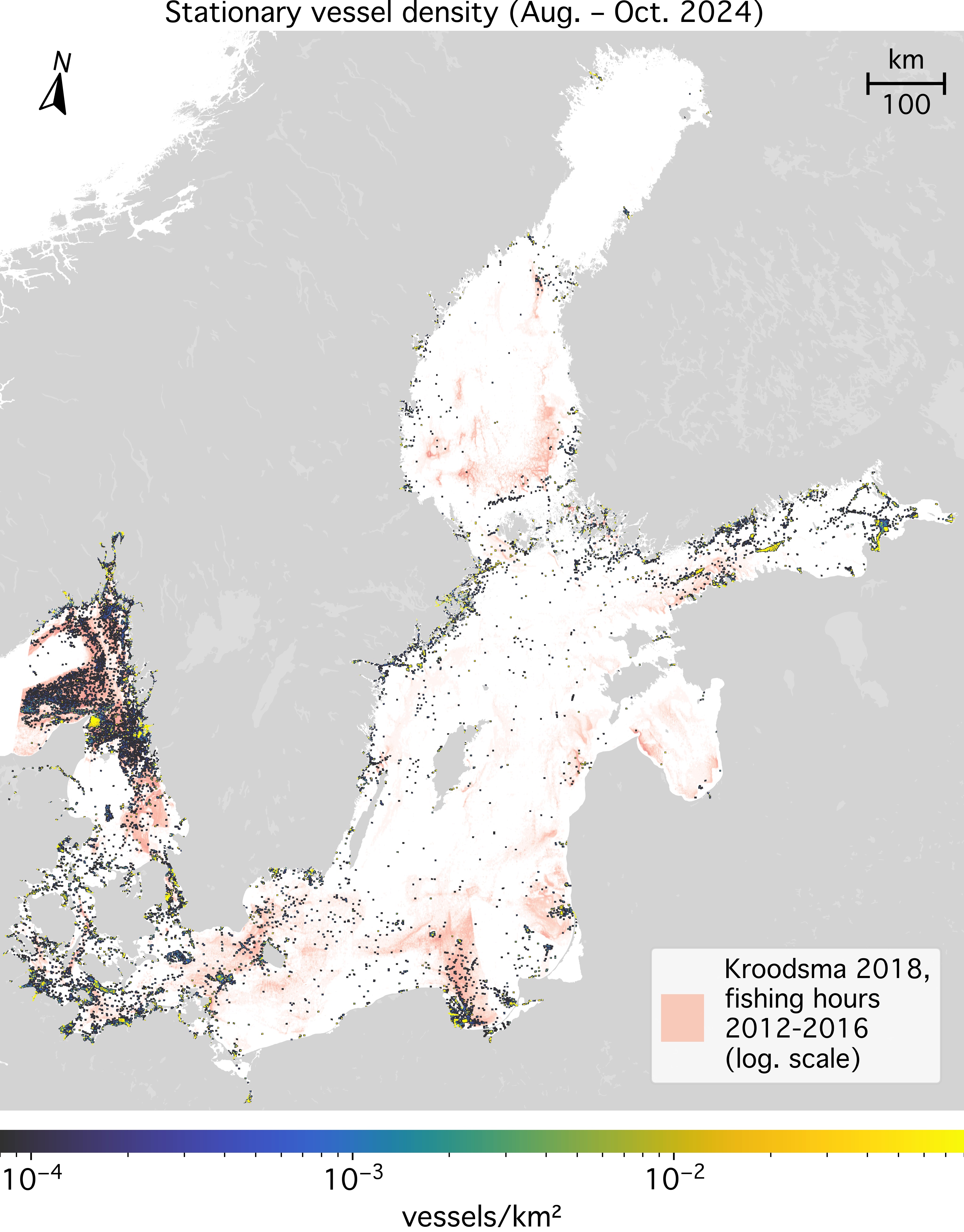}}
    \caption{Vessel density in the Baltic Sea, average from August to October 2024. Densities are shown for all vessel types combined. The~left panel also shows inferred coastal port areas (Section~\ref{subsec:results_ports}), with~the busiest (Table~\ref{tab:busiest_ports}) marked in green, and~additional large ports (Table~\ref{tab:biggest_ports}) in~yellow. (\textbf{a}) Combined density of all moving and stationary vessels (55.2\% of the average 4061 simultaneous vessels shown). (\textbf{b})~Stationary vessels with positions found at \SI{1}{\km} accuracy (48.6\% of 3287 vessels shown). Fishing data from \citet{Kroodsma2018}.}
    \label{fig:density_map_baltic}
\end{figure}

\Cref{fig:density_map_baltic} presents the spatial distribution of average vessel counts for all vessels (Figure~\ref{fig:density_map_baltic}a) and for the subset of stationary vessels (Figure~\ref{fig:density_map_baltic}b). Because~moving vessels are not represented on incorrectly reconstructed tracklets, and~stationary vessels are omitted if their position cannot be determined with at least \SI{1}{\km} accuracy (Section~\ref{subsec:maps}), these figures represent only 55\% of all observed vessels and 49\% of stationary vessels. 

Figure~\ref{fig:density_map_baltic}a shows the limitations of our AIS dataset in accurately capturing vessel activity in the Gulf of Riga and Bothnia. Additionally, large distances to the next receivers between Gotland and the Latvian coast result in poorly reconstructed tracklets. This is indicated by the seemingly high vessel density close to the southeast coast of Gotland, which is an artifact of the incomplete data. In~Figure~\ref{fig:density_map_baltic}b, a~large number of stationary vessels can be seen in the northern Kattegat and Skagerrak regions. This aligns with the high fishing activity in these areas, as~indicated by the red-shaded area \mbox{from \citet{Kroodsma2018}}; see also \citet{HELCOM2023b} and the reproduced result from \citet{Paolo2024} in Figure~\ref{fig:tracked_vessels}b.

\unskip

\subsection{Traffic Within, into, and~from the Baltic~Sea}
\label{subsec:results_traffic}

\begin{figure}[t!]
\centering
    \subfloat[\centering\label{fig:vessel_flux_baltic}]{\includegraphics[width=0.49\textwidth]{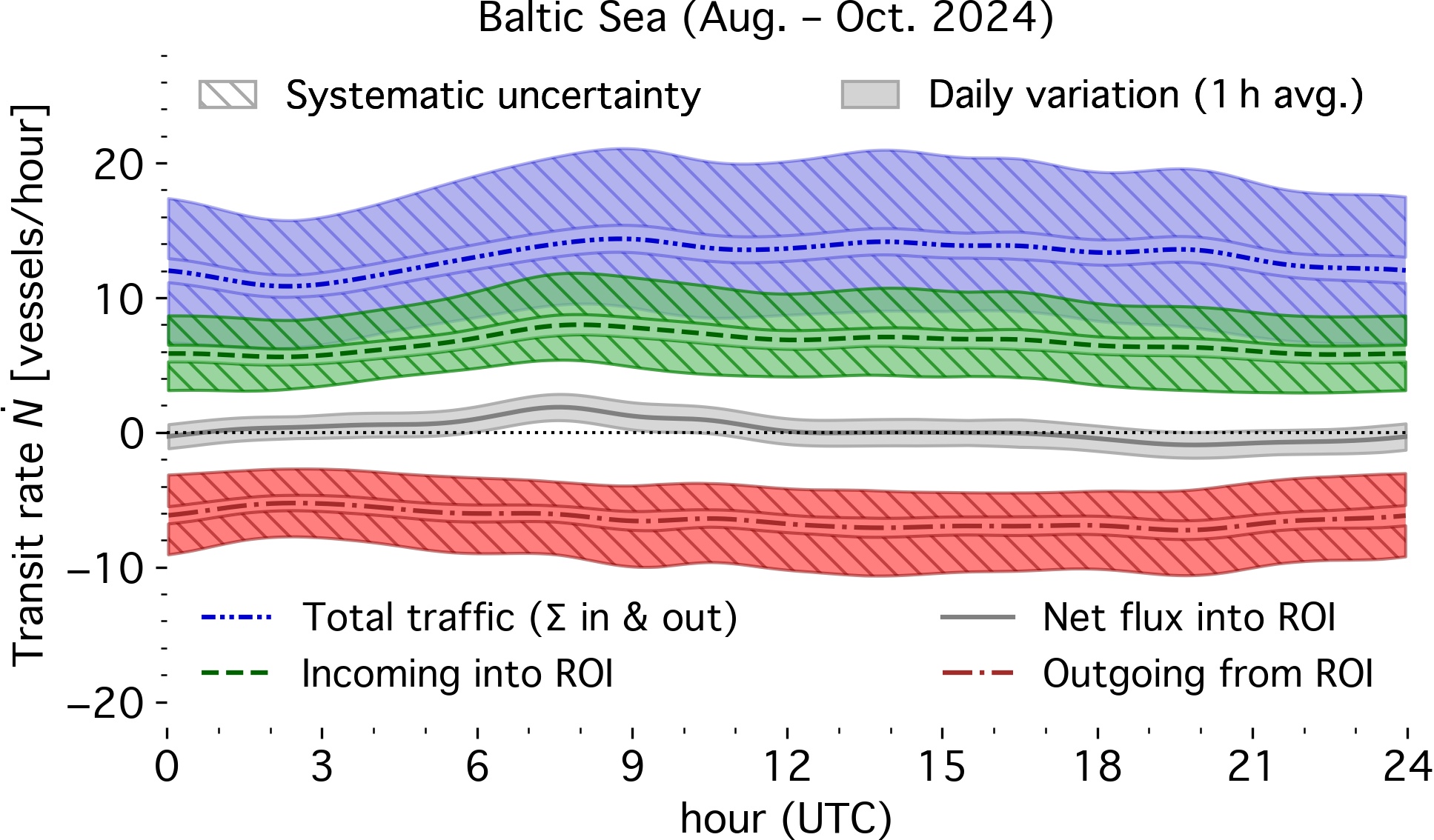}}\\\vspace{0.1cm}
    \subfloat[\centering\label{fig:vessel_flux_skagerrak}]{\includegraphics[width=0.49\textwidth]{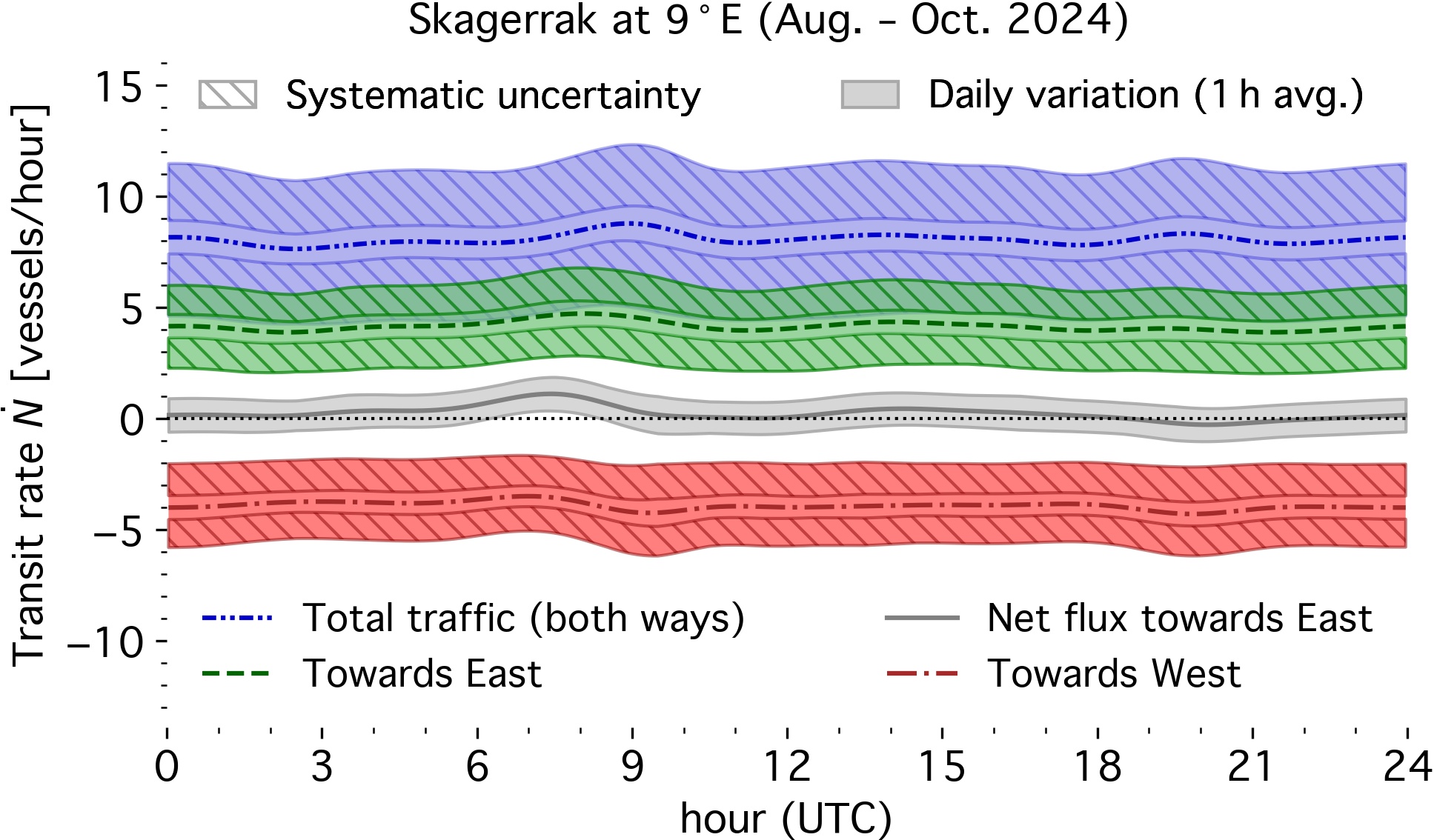}}\hfill
    \subfloat[\centering\label{fig:vessel_flux_kiel}]{\includegraphics[width=0.49\textwidth]{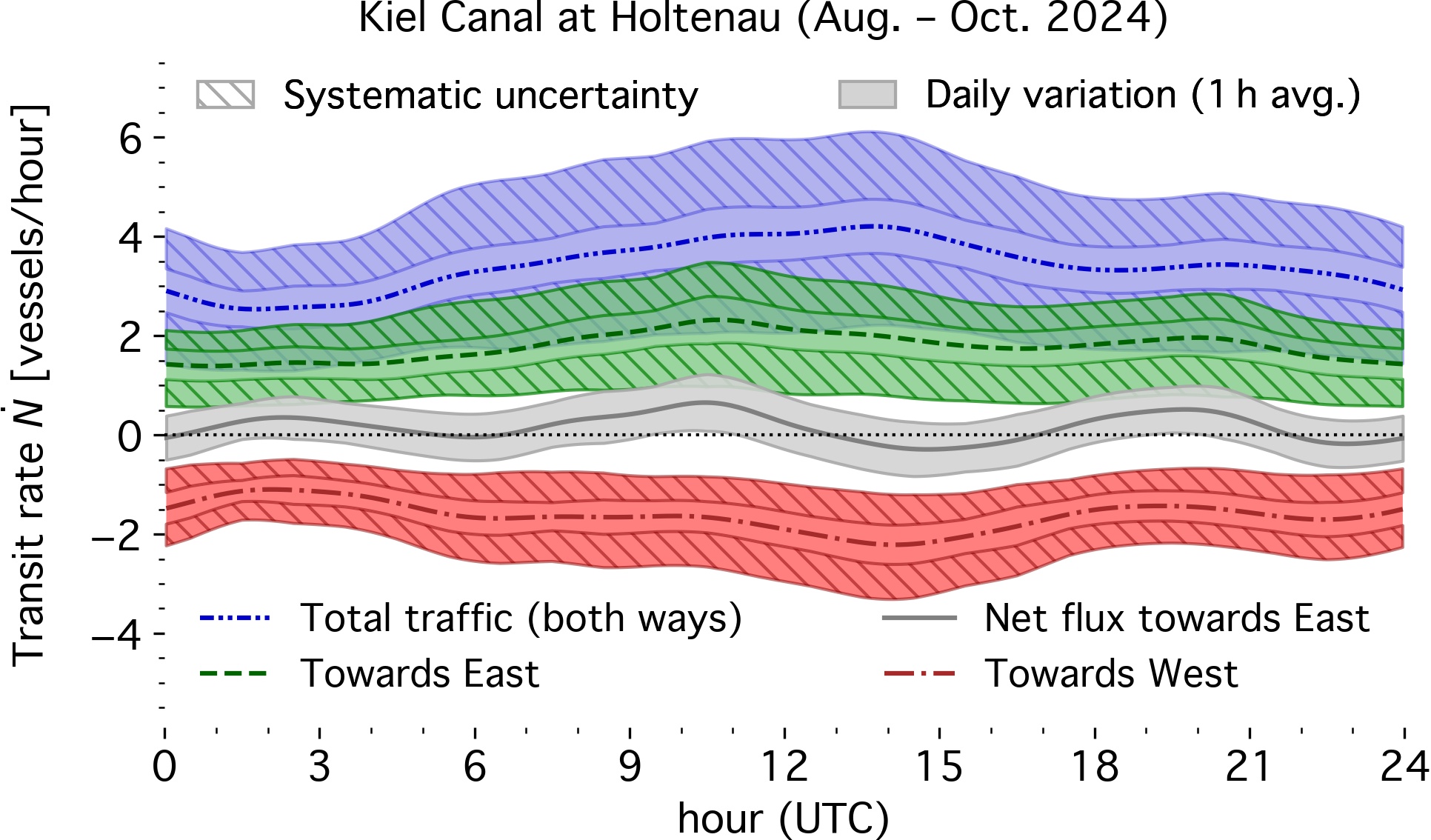}}
    \caption{\textls[-15]{Vessels per hour entering or leaving the Baltic Sea ROI through all transit areas (\textbf{top panel}), only the Skagerrak Strait (\textbf{bottom left panel}), or~the Kiel Canal (\textbf{bottom right panel}). 
    No systematic uncertainty is shown for the net flux for simplicity. (\textbf{a})  Vessels entering or leaving the Baltic Sea during the day. Shown are net traffic (gray curve and band without systematic uncertainties); 
    inbound (green dashed curve and bands), outbound (red dash-dotted curve and bands), and~total traffic crossing the ROI boundary (blue dash-double-dotted curve and bands). (\textbf{b}) Part of the traffic shown in \Cref{fig:vessel_flux} passes through the Skagerrak. 
    (\textbf{c}) Part of the traffic shown in \Cref{fig:vessel_flux}a passes through the Kiel Canal. 
    In the bottom panels,
    a~course direction towards the east corresponds to entering vessels, and~towards the west to leaving vessels.}}
    \label{fig:vessel_flux}

    \vspace{20pt}  

	\includegraphics[width=0.55\textwidth]{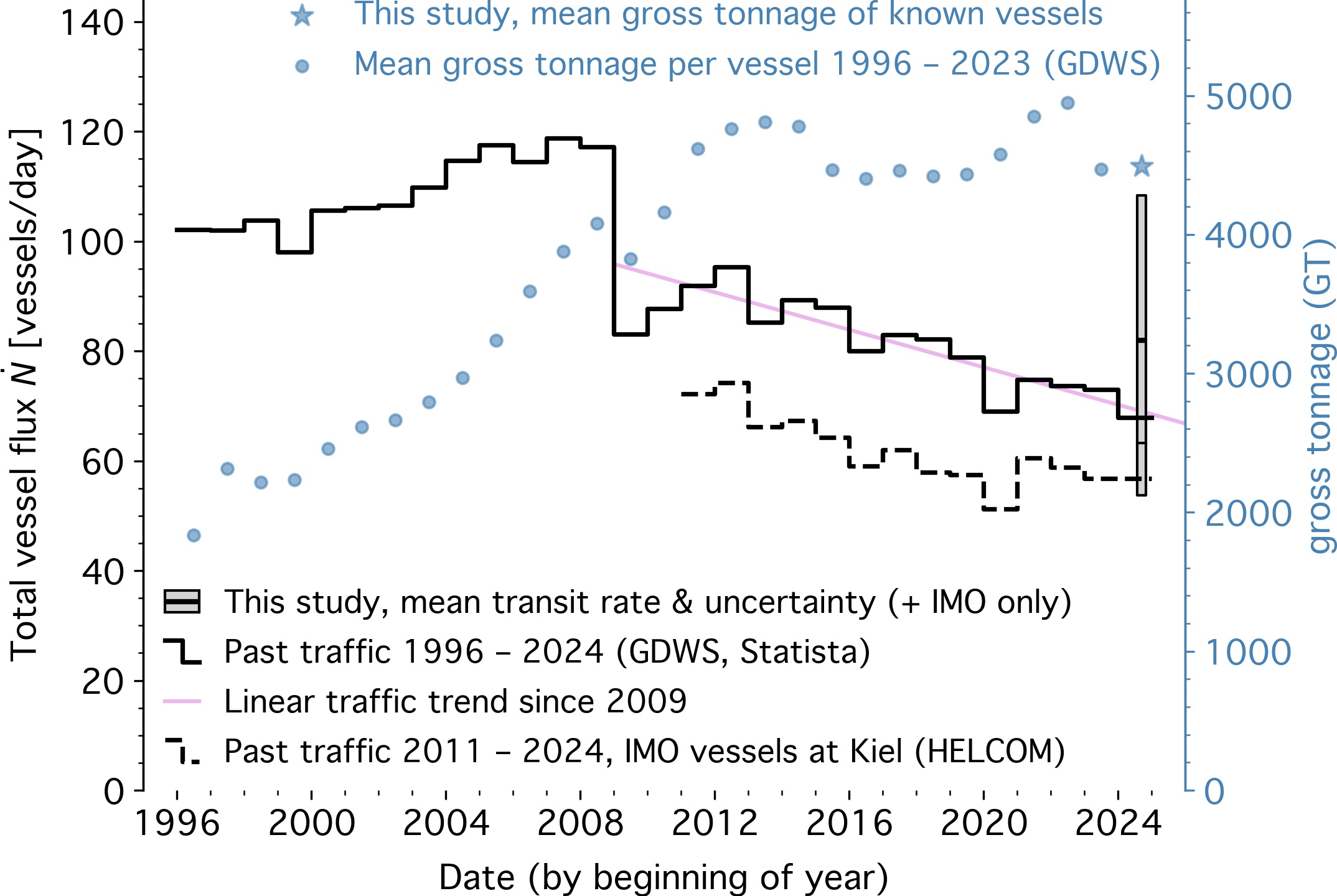} 
		  \caption{
Timeline of historic traffic (full and section passages; left black scale) and average GT per vessel (right blue scale) through the Kiel Canal, compared with our results. For~our results, the~gray box marks the uncertainty range of the transit rate into/from the Canal, with~the thick horizontal line the mean all-vessel transit rate, and~the thin horizontal line the rate for IMO vessels only. Our mean estimated GT is marked by the blue star. Results are compared against the data from \citep{Statista2019,Statista2024a,GDWS2024,GDWS2025} (black curve and blue dots) and \citet{HELCOM2023a} (black dashed curve).
      }
      \label{fig:vessels_historic_kiel}
\end{figure}

\Cref{fig:vessel_flux}a--c resolve the overall transit rate and the rates through the most prominent transit areas over the course of the day. We observe a slight net inflow of vessels into the Baltic Sea in the morning and net outflow in the afternoon, primarily driven by traffic through the Skagerrak (black solid lines in Figure~\ref{fig:vessel_flux}a,b). Notably, we observe an approximately eight-hour period of net traffic direction change at Kiel Holtenau (black solid line in Figure~\ref{fig:vessel_flux}c). 
The transit rates through these areas are listed in \Cref{tab:vessel_numbers_flux}. 
\begin{table}[H]
\footnotesize
\caption{Transit 
 rates  $\dot{N}$: Vessels entering or leaving the Baltic Sea each day (August--October 2024).}\label{tab:vessel_numbers_flux}
\begin{adjustwidth}{-1.5cm}{0cm}
\newcolumntype{C}{>{\centering\arraybackslash}X}
\begin{tabularx}{\fulllength}{lccCCCC}
\toprule 
\textbf{Vessel Type}& \multicolumn{2}{c}{\textbf{All Transit Areas}} & \multicolumn{2}{c}{\textbf{Skagerrak at $\boldsymbol{9^\circ}$E}} & \multicolumn{2}{c}{\textbf{Kiel Canal at Holtenau}}\\
\midrule
 All vessels & 
  $313 \pm 23\,{}^{+121}_{-\;\,98}$ & $100\%\;$     & 
  $194 \pm 14\,{}^{+ 61}_{-    67}$ & $100\%\;$     & 
  $ 82 \pm  6\,{}^{+ 26}_{-    28}$ & $100\%\;\;\,$ \\ 
\midrule

Passenger, high-speed & 
 $\;\,32 \pm \;\,2\,{}^{+\;\,17}_{-\;\,31}$ & $   10\%$ &
 $\;\,17 \pm \;\,1\,{}^{+ \;\,3}_{- \;\,1}$ & $\;\,9\%$ &
 $\quad\,0.7\pm0.05\,{}^{+0.2}_{-0.7}$      & $\;\,1\%$ \\[0.15cm]
 
Law enforcement,  military & 
 $     \;3.0\pm 0.2\,{}^{+1.2}_{-1.8}$ & $\;\,1\%$ &
 $ \;\;\,1.7\pm 0.1\,{}^{+0.5}_{-0.8}$ & $\;\,1\%$ &
 $\quad\,0.4\pm0.03\,{}^{+0.1}_{-0.4}$ & <$1\%$    \\[0.15cm]

Cargo & 
 $   144\pm    11\,{}^{+\;\,44}_{-\;\,65}$ & $46\%$ &
 $\;\,97\pm \;\,7\,{}^{+    21}_{-    34}$ & $50\%$ &
 $\;\,34\pm \;\,3\,{}^{+ \;\,5}_{    -28}$ & $41\%$ \\[0.15cm]
 
Pilot, tug, rescue, diving/dredging &  
 $   \;\,37\pm \;\,3\,{}^{+\;\,18}_{-\;\,29}$ & $12\%$    &
 $\;\;\,3.0\pm   0.2\,{}^{+   0.8}_{-   1.4}$ & $\;\,2\%$ &
 $   \;\,33\pm \;\,2\,{}^{+    14}_{-    28}$ & $40\%$    \\[0.15cm]
 
Tanker & 
 $   \;\,57\pm \;\,4\,{}^{+\;\,15}_{-\;\,21}$ & $18\%$ &
 $   \;\,46\pm \;\,3\,{}^{+ \;\,7}_{-    16}$ & $24\%$ &
 $\;\;\,7.9\pm   0.6\,{}^{+   0.3}_{-   4.8}$ & $10\%$ \\[0.15cm]
 
Others including fishing & 
 $ \;\,39\pm \;\,3\,{}^{+\;\,50}_{-\;\,17}$ & $     12\%$ &
 $ \;\,29\pm \;\,2\,{}^{+    37}_{-    14}$ & $     15\%$ &
 $\;\,6.2\pm   0.4\,{}^{+ \;\,8}_{- \;\,3}$ & $\;\,8\%\;$ \\
 \midrule
   
IMO vessels &
 $   266\pm    19\,{}^{+123}_{-\;\,99}$ & $85\%$ & 
 $   176\pm    13\,{}^{+ 59}_{-    59}$ & $91\%$ &
 $\;\,63\pm \;\,5\,{}^{+ 33}_{-    37}$ & $77\%$ \\[0.2cm]
  
Vessels with GT $<10{,}000$ &
 $   173\pm    12\,{}^{+110}_{-\;\,73}$ & $\geq$$55\%\;\;\;$ &
 $\;\,92\pm \;\,7\,{}^{+ 55}_{-    37}$ & $\geq$$47\%\;\;\;$ &
 $\;\;57\pm \;\,4\,{}^{+ 34}_{-    34}$ & $\geq$$70\%\;\;\;$ \\[0.15cm]
 
Vessels with GT $\geq10{,}000$ &
 $   \;\,95\pm \;\,7\,{}^{+\;\,26}_{-\;\,26} $ & $\geq$$30\%\;\;\;$   &
 $   \;\,82\pm \;\,6\,{}^{+    12}_{-    22} $ & $\geq$$42\%\;\;\;$   &
 $\;\;\,8.6\pm0.6\,{}^{   +0.5   }_{-   2.9} $ & $\geq$$10\%\;\;\;\;$ \\
\bottomrule
\end{tabularx}
\end{adjustwidth}
\end{table}

\Cref{fig:vessels_historic_kiel} compares our results of traffic into or from the Kiel Canal at the Holtenau lock with historic data from other sources. The~Kiel Canal is the world's busiest artificial waterway in terms of passages \citep{Heitmann2013}. On~average, around 80 vessels have traveled each day through the full canal or a section of it over the last decade (black solid line based on data from~\citep{Statista2019,Statista2024a,GDWS2024,GDWS2025}). However, traffic has declined since 2009 (violet line). Our 2024 estimate (gray box) is consistent with previous years and the overall trend within uncertainties. Our mean all-vessel transit rate (thick black horizontal line inside the box) is 20\% larger than the reported all-year traffic in 2024. While the historic reference data in principle include sectional passages, it may not account for all short-term trips from or into the canal by pilot, tug, or~dredging vessels, which are included in our analysis. Additionally, Figure~\ref{fig:vessels_historic_kiel} displays the vessel transits at Holtenau as inferred from the HELCOM data \citep{HELCOM2023a} (black dashed curve), which can be compared with our result for IMO vessels only (thin black horizontal line inside the gray box). Here, too, our result is slightly (11\%) 
higher than the 
all-year data from 2024.

While fewer vessels have passed the Kiel Canal over the last decades, the~average vessel gross tonnage on full and sectional passages increased in the years prior to 2023, but~dropped in 2023 (blue dots in Figure~\ref{fig:vessels_historic_kiel}; no
 all-passage GT values have been officially published for 2024).
Based on the available gross tonnage information for IMO vessels, our three-month window suggests a value lower than before 2023, also in 2024 (GT$\;=4493$, blue star in Figure~\ref{fig:vessels_historic_kiel}). This indicates that not only has the number of vessels decreased, a~long-term trend towards fewer, but~larger vessels since the 1960s \citep{GDWS2024}, but~also the total volume of goods transported through the Kiel Canal has recently declined. Interestingly, such a net decline in transported volume may also be connected to the trend towards larger vessels, which may exceed the capacity of the Kiel Canal and are, therefore, routed via the~Skagerrak.

\subsection{Port Areas in the Baltic~Sea}
\label{subsec:results_ports}

Using a map segmentation of the vessel density (Section~\ref{subsec:ports}), we find 145 overdensities in the ROI with \SI{3.9}{\square\km}, \SI{5.6}{\square\km}, and~\SI{33.0}{\square\km} in median, mean, and~maximum size, respectively. In total, 95\% of those areas have a barycenter location within \SI{2.7}{\km} from the coast. Among~these coastal port areas, we identify 44\% of 143 Baltic Sea ports in the World Port Index (WPI, \citep{WPI2025}) classified as small or larger and 85\% of the large ports, with~Riga, Sillamae, and~Oulu missing among the latter. Conversely, 67\% of the identified overdensities correspond to a WPI-listed port. 
Our algorithm also detects several short-distance ferry connections without major port infrastructure. \Cref{fig:density_map_baltic}a shows the port areas located within \SI{2.7}{\km} of the~shoreline. 

\Cref{tab:busiest_ports} lists the ten busiest identified ports in terms of arrivals. These ports include the following pairs of ferry lines in the Baltic Sea with the highest frequency: the Helsingør–Helsingborg ferry between Denmark in Sweden serving every 20 min during daytime (first two rows), the~Moss–Horten ferry in the Oslo Fjord (Norway; forth and eighth rows), and~the Rødby–Puttgarden connection between Denmark and Germany (seventh and ninth rows). For~all these ports, except~Moss, we obtain an arrival rate equal to or greater than expected from the corresponding ferry timetables. For~Moss, we inferred only 53.7 arrivals per day, while we expect 54 arrivals from the ferry connection alone. Our arrival rates are about a factor of ten larger than the daily AIS-inferred port calls published by \citet{Arslanalp2021}.
\Cref{tab:biggest_ports} ranks the ten largest ports by the number of vessels moored in the areas found in our analysis. We find that Gothenburg is the largest port, also by area, with~almost two vessels per \SI{}{\square\km} on an area larger than \SI{30}{\square\km}. The~remaining ports in the table belong to major urban areas in the Baltic Sea region, among~them six of the ten ports found by \citet{Synak2023} with the largest cargo turnover in the first half of 2023: Gda\'nsk, Gothenburg, Szczecin, Rostock, Klaip\'eda, and~Gdynia. 

We also list the most common destination names provided by the AIS in these tables. No standard format exists for these data. Sometimes, the~arrival port is provided either by UN/LOCODE or real name, sometimes both the last departure and destination ports are provided, and~sometimes the field is used to provide information about the carrier. Additionally, spelling mistakes were frequently observed in our data. This renders, in particular, the human-inputted AIS information challenging for automated data analysis \citep{Harati-Mokhtari2007,Emmens2021,Sun2025}.

\begin{table}[t]  
\caption{The ten
 busiest port areas in the Baltic Sea (August--October 2024) in terms of arrivals. }
\label{tab:busiest_ports}
\footnotesize
\renewcommand{\arraystretch}{1.2}
\setlength{\tabcolsep}{1mm} 
\begin{adjustwidth}{-1.5cm}{0cm}
\setlength{\cellWidtha}{\fulllength/9-2\tabcolsep-.40in}
\setlength{\cellWidthb}{\fulllength/9-2\tabcolsep-.40in}
\setlength{\cellWidthc}{\fulllength/9-2\tabcolsep-.20in}
\setlength{\cellWidthd}{\fulllength/9-2\tabcolsep+.40in}
\setlength{\cellWidthe}{\fulllength/9-2\tabcolsep+.60in}
\setlength{\cellWidthf}{\fulllength/9-2\tabcolsep-0in}
\setlength{\cellWidthg}{\fulllength/9-2\tabcolsep-0in}
\setlength{\cellWidthh}{\fulllength/9-2\tabcolsep-0in}
\setlength{\cellWidthi}{\fulllength/9-2\tabcolsep-0in}
\scalebox{1}[1]{\begin{tabularx}{\fulllength}{>{\centering\arraybackslash}m{\cellWidtha}>{\centering\arraybackslash}m{\cellWidthb}>{\centering\arraybackslash}m{\cellWidthc}>{\centering\arraybackslash}m{\cellWidthd}>{\centering\arraybackslash}m{\cellWidthe}>{\centering\arraybackslash}m{\cellWidthf}>{\centering\arraybackslash}m{\cellWidthf}>{\centering\arraybackslash}m{\cellWidthf}>{\centering\arraybackslash}m{\cellWidthf}}

\toprule
\textbf{Rank} & 
   $\boldsymbol{\varphi}$ \textbf{[}$\boldsymbol{^\circ}$\textbf{]}& 
   $\boldsymbol{\lambda}$ \textbf{[}$\boldsymbol{^\circ}$\textbf{]}& 
  \textbf{Name} & 
 \textbf{Most Common AIS} \textbf{Destination Name}  &
  \textbf{Arrivals per Day}& 
  \textbf{Vessels in } \linebreak  \textbf{Port}&
  \textbf{Port Area} \textbf{[km\textsuperscript{2}]}&
  \textbf{Density} \linebreak  \textbf{[\mbox{Vessels}/km\textsuperscript{2}]}
  \\
\midrule
1 &$56.040$ & $12.679$ &
 Helsingborg (SE) &
 FERRY DK &
$74.4 $ &
$10.0$ &
$\;\,8.9$ &
$1.1$
 \\
2& $56.035$ & $ 	12.612$ &
 Helsing\o{}r (DK) &
 FERRY DK &
 $71.9$ &
 $\;\,6.8$ &
 $\;\,7.7$ &
 $0.9$
\\
3& $57.673$ & $11.854$ &
 Gothenburg (SE) &
 STYRSOBOLAGET FRAKT&
 $68.8$ &
 $55.8$ &
 $33.0$ &
 $1.7$
\\
4& $59.419$ & $10.479$ &
 Horten (NO) &
 HORTEN MOSS &
 $63.8$ &
 $12.1$ &
 $\;\,9.0$ &
 $1.3$
 \\
5& $59.898$ & $10.729$
 &
 Oslo (NO)&
 NOOSL&
 $62.0$ &
 $18.7$ &
 $\;\,9.5$ &
 $2.0$
 \\
6 & $54.644$ & $11.354$ &
 R\o{}dbyhavn (DK) &
 DKROD & 
 $57.0$ &
 $20.3$ &
 $17.0$ &
 $1.2$
 \\
7 & $59.419$ & $10.638$ &
 Moss (NO) &
 HORTEN MOSS& 
 $53.7$ &
 $\;\,4.3$ & 
 $\;\,4.4$ &
 $1.0$
 \\
8&  $54.502$ & $11.229$ &
 Puttgarden (DE) &
 DKROD & 
 $47.4$&
 $\;\,6.4$ &
 $\;\,6.3$ &
 $1.0$
 \\
 9 &$57.610$ & $11.788$ &
 Southern~Gothenburg archipelago (SE) &
 STYRSOBOLAGET FRAKT  &
 $37.4$ &
 $\;\,6.3$ &
 $\;\,6.9$ &
 $0.9$
 \\
10& $59.319$ & $18.104$ &
 Stockholm (SE) &
 SL LINJE &
 $35.0$ &
 $48.4$ &
 $\;\,8.1$ &
 $5.9$
 \\
\bottomrule
\end{tabularx}}
\end{adjustwidth}
\end{table}

\begin{table}[t] 
\caption{The ten 
 largest port areas in the Baltic Sea (August--October 2024) in terms of average number of vessels in port. }
\label{tab:biggest_ports}
\footnotesize
\renewcommand{\arraystretch}{1.2}
\setlength{\tabcolsep}{1mm} 
\begin{adjustwidth}{-1.5cm}{0cm}
\setlength{\cellWidtha}{\fulllength/9-2\tabcolsep-.40in}
\setlength{\cellWidthb}{\fulllength/9-2\tabcolsep-.40in}
\setlength{\cellWidthc}{\fulllength/9-2\tabcolsep-.20in}
\setlength{\cellWidthd}{\fulllength/9-2\tabcolsep+.20in}
\setlength{\cellWidthe}{\fulllength/9-2\tabcolsep+.70in}
\setlength{\cellWidthf}{\fulllength/9-2\tabcolsep-0in}
\setlength{\cellWidthg}{\fulllength/9-2\tabcolsep-0in}
\setlength{\cellWidthh}{\fulllength/9-2\tabcolsep-0in}
\setlength{\cellWidthi}{\fulllength/9-2\tabcolsep+.10in}
\scalebox{1}[1]{\begin{tabularx}{\fulllength}{>{\centering\arraybackslash}m{\cellWidtha}>{\centering\arraybackslash}m{\cellWidthb}>{\centering\arraybackslash}m{\cellWidthc}>{\centering\arraybackslash}m{\cellWidthd}>{\centering\arraybackslash}m{\cellWidthe}>{\centering\arraybackslash}m{\cellWidthf}>{\centering\arraybackslash}m{\cellWidthf}>{\centering\arraybackslash}m{\cellWidthf}>{\centering\arraybackslash}m{\cellWidthf}}

\toprule
\textbf{Rank} & 
   $\boldsymbol{\varphi}$ \textbf{[}$\boldsymbol{^\circ}$\textbf{]}& 
   $\boldsymbol{\lambda}$ \textbf{[}$\boldsymbol{^\circ}$\textbf{]}& 
  \textbf{Name} & 
 \textbf{Most Common AIS} \textbf{Destination Name}  &
  \textbf{Arrivals per Day}& 
  \textbf{Vessels in} \linebreak   \textbf{Port}&
  \textbf{Port Area} \textbf{[km\textsuperscript{2}]}&
  \textbf{Density} \textbf{[\mbox{Vessels}/km\textsuperscript{2}]}
  \\
\midrule
1 & $57.673$ & $11.854$ &
 Gothenburg (SE) &
 STYRSOBOLAGET FRAKT&
 $68.8$ &
 $55.8$ &
 $33.0$ &
 $1.7$
 \\
2 & $59.319$ & $18.104$ &
 Stockholm (SE) &
 SL LINJE &
 $35.0$ &
 $48.4$ &
 $\;\,8.1$ &
 $5.9$
\\
3& $54.390$ &
 $18.671$ &
 Gda\'nsk (PL) &
 GDANSK \mbox{ANCHORAGE} & 
 $18.8$ &
 $43.5$ &
 $23.6$ &
 $1.8$
\\
4& $55.681$ & $21.129$ &
 Klaip\'eda (LT)&
 KLAIPEDA LITHUANIA &
 $14.3$ &
 $41.5$ &
 $11.4$ &
 $3.6$
 \\
5&$54.531 	$ & $18.546$ &
 Gdynia (PL) &
 GDYNIA POLAND & 
 $14.3$ & 
 $40.7$ &
 $18.8$ &
 $2.2$
 \\
6 &$54.177$ & $12.096$ &
 Rostock (DE) &
 DERSK \mbox{HANSESAIL} &
 $33.9$ &
 $38.9$ &
 $17.4$ &
 $2.2$
 \\

7& $54.348$ & $10.154$ &
 Kiel Fjord without $\quad$ Laboe (DE) &
 KIEL PILOT & 
 $32.1$ & 
 $37.1$ &
 $16.1$ &
 $2.3$
 \\
8&  $53.444$ & $14.588$ &
 Szczecin (PL) &
 SZCZECIN PLSZZ &
 $\;\,9.5$ &
 $29.2$ &
 $\;\,9.5$ & 
 $3.1$
 \\
9&$55.694$ & $12.621$ &
 Kopenhagen (DK) &
 DKCPH & 
 $13.4$ &
 $23.8$ &
 $23.8$ &
 $1.0$
 \\
10&  $57.694$ & $11.662$ &
 Northern Gothenburg archipelago (SE) &
 BJORKO & 
 $\;\,7.8$ &
 $22.8$ &
 $14.1$ &
 $1.6$
 \\
\bottomrule
\end{tabularx}}
\end{adjustwidth}
\end{table}

\begin{figure}[t]
\centering
    \subfloat[\centering\label{fig:kiel_density}]{
        \includegraphics[width=0.24\textwidth]{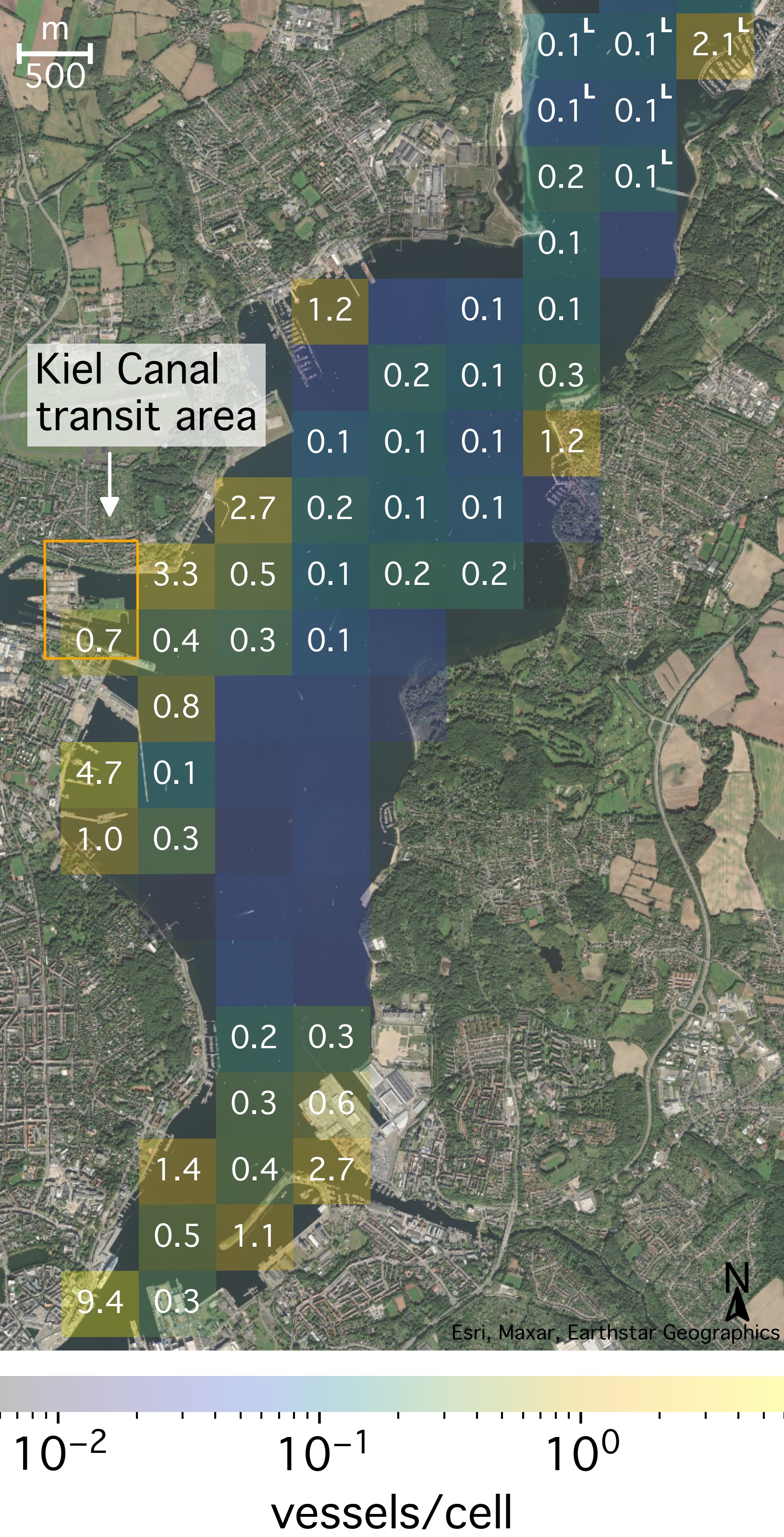}
    }
    \hspace{0.2cm}
    \subfloat[\centering\label{fig:kiel_crossing}]{
        \includegraphics[width=0.24\textwidth]{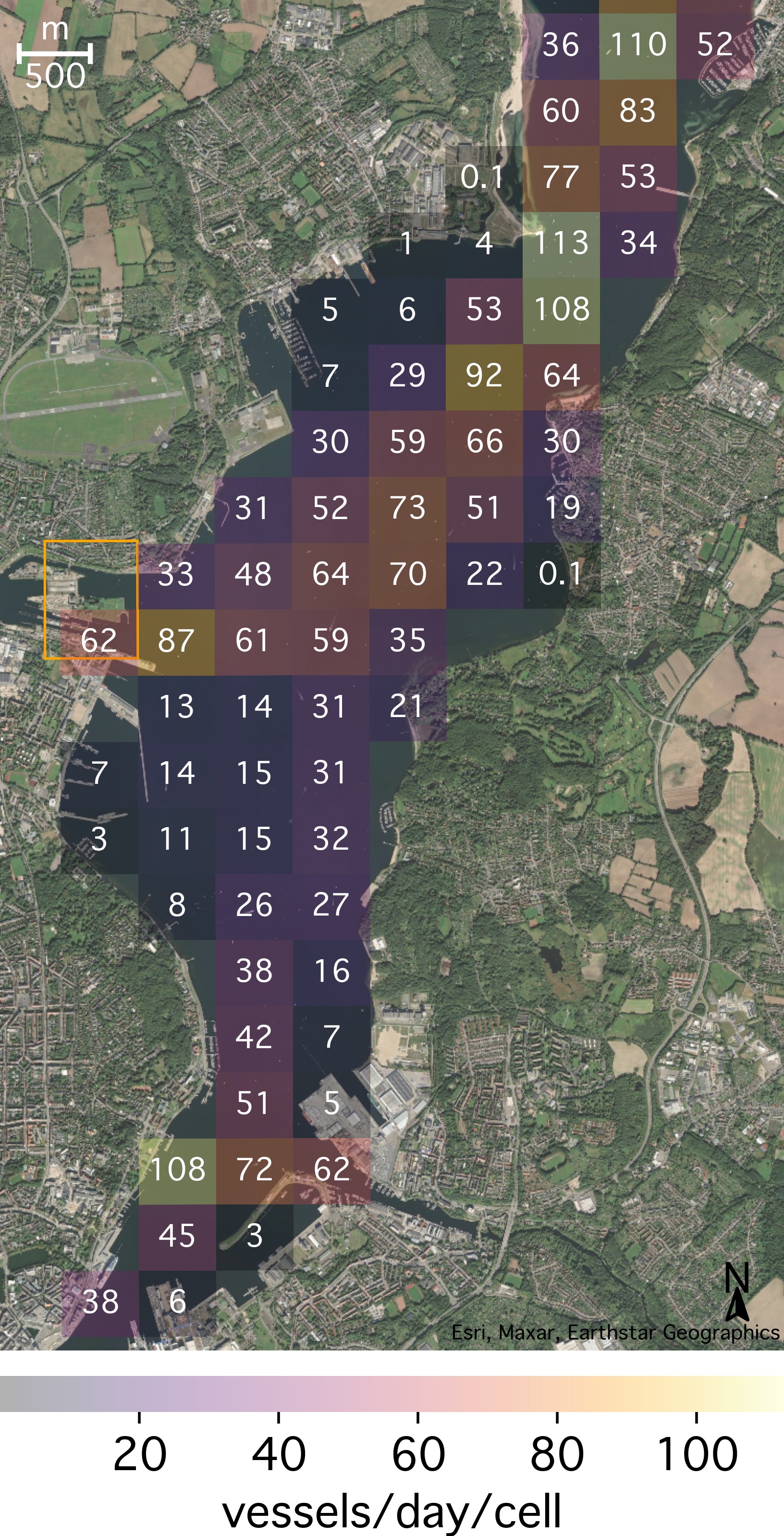}
    }
    \caption{\textls[-15]{Vessel density and traffic in the port areas of the Kiel Fjord (Kieler F\"orde) 
    and part of Laboe (marked by an L in the left panel). No densities are indicated for less than $0.05$ vessels per~cell. (\textbf{a})~ Average vessel density of moving and stationary vessels. (\textbf{b}) Daily cell crossings by moving~vessels.}}
    \label{fig:kiel_closeup}
\end{figure}

Finally, \Cref{fig:kiel_closeup} zooms into the Kiel Fjord port area (ranked 7 in Table~\ref{tab:biggest_ports}), showing spatially resolved vessel density (Figure~\ref{fig:kiel_closeup}a) and traffic in terms of cell crossings (\mbox{Figure~\ref{fig:kiel_closeup}b}). This port area also contains the Kiel Canal transit area, marked by the orange box. Cells belonging to the separate area of Laboe port are marked by the letter `L' in Figure~\ref{fig:kiel_closeup}a. We verified the average densities (Figure~\ref{fig:kiel_closeup}a) by comparing against momentary live data from \citet{VesselFinder2025} and found consistent~results. 

\section{Discussion}
\label{sec:discussion}

In this work, we have presented a method to infer maritime activities from open-access
AIS positioning data. By~analyzing three months of vessel activity in the Baltic Sea, we have shown that available open data is of sufficient quality to study coastal regions as large as \SI{1000}{\km} in size, and~provide insights that are competitive with those derived from proprietary data. Our estimates of both the average vessel count in the area and of the traffic into and from the Kiel Canal agree within 20\% with previous studies, after~we have assessed multiple sources of systematic uncertainty and have justified their quantification. Using a simple stop-identification algorithm and density thresholding, we have identified half of the 143 major WPI-listed ports in the region, and we obtained metrics for these port areas consistent with other sources. This shows that reliable statistics on coastal-water vessel activity can be obtained from publicly available and community data, supporting research in economic, environmental, and~transportation~sciences.

A major challenge in the limited time-period analysis of regional data is the treatment of edge effects in space and time. At~the spatial boundaries, this involves properly identifying transiting or idle vessels. If~data is available, expanding the analysis region beyond the actual ROI can mitigate this; for example, to~infer the Kiel Canal traffic with larger precision, as presented in this analysis. Also, if~training data is provided, machine-learning methods are well suited for such a discrimination task. However, extended or reference data may not always be available. In~a separate work, we will show that the rule-based cleansing and discrimination techniques presented in this paper can also be robustly applied at higher spatial resolution to AIS data only covering single port areas and from single receivers with fluctuating and fading coverage. Temporally, we have found the chosen three-month interval for the given ROI to be considerably affected by edge effects. For~longer periods or continuous analysis, techniques like split--apply--combine strategies~\citep{Wickham2011} may be indispensable to manage memory and to efficiently parallelize~computation. 

In this study, we did not address the detection of intentionally manipulated AIS data. Vessels that temporarily deactivate their transceivers for illegal activities are not distinguished from those omitting position signals unintentionally \citep{Harati-Mokhtari2007,Emmens2021} or due to other data gaps. During~such gaps, vessel routes are always reconstructed as the most plausible, valid shortest paths or interpreted as mooring states. Revealing the true positions of temporarily dark vessels, or~of those not using AIS at all, requires complementary data sources such as satellite imagery \citep{Paolo2024}. Nevertheless, several approaches have been proposed to detect voluntary transmission gaps in AIS data streams alone \citep{Sharma2024,Bernabe2024}. The~same applies to identifying and removing intentionally injected but otherwise plausible false AIS messages~\citep{Androjna2023,Kessler2024,Louart2024}.
Further research is needed to determine whether these detection methods can be reliably applied to community data and whether open-access AIS data  can be effectively protected against spoofing attacks. Additional improvements to the analysis include incorporating AIS-B data and, in general, applying machine-learning techniques for error correction of false or patchy data. This is left for future~work.

\vspace{6pt} 


\funding{This research received no external~funding.}

\dataavailability{The original data presented in the study and research outputs are openly available at \url{https://dx.doi.org/10.6084/m9.figshare.29062715}. 
}

\acknowledgments{\textls[-15]{This work relied on the 
 GeographicLib \citep{Karney2013} and n-vector \citep{Gade2010} software for geodesic calculations, SciPy \citep{Virtanen2020}, scikit-image, GDAL, QGIS; and the Basemap, Colorcet, cmocean, Palletable, and Font Awesome packages for visualization. 
The author would like to \mbox{thank \citet{aisstream2025}} and the network's receiver stations for providing the AIS data, \mbox{and~\citet{VesselFinder2025}} for the kind permission to use their gross tonnage data. Michael Dziomba, Henrik Hanssen, Marcel Strzys, Hiroki Tsuda, Thomas Zwetyenga, and the two anonymous reviewers provided valuable comments that helped to improve the quality of the manuscript.}}

\conflictsofinterest{The author declares no conflicts of~interest. Author Moritz Hütten is employed by the company GRID Inc. The author declares that the research was conducted in the absence of any commercial or financial relationships that could be construed as a potential conflict of interest. GRID Inc. approved this manuscript for publication; however, it had no role in the design of the study; in the collection, analysis, or interpretation of data; in the writing of the manuscript; or in the decision to publish the results.
} 


\abbreviations{Abbreviations}{
The following abbreviations are used in this manuscript:
\\

\noindent 
\begin{tabular}{@{}ll}
AIS & Automatic Identification System\\
GT & Gross Tonnage\\
HELCOM & Baltic Marine Environment Protection Commission (Helsinki Commission)\\
IMO & International Maritime Organization\\
MMSI & Maritime Mobile Service Identity (number)\\
RDP & Ramer-Douglas-Peucker (simplification algorithm)\\
ROI & Region Of Interest\\
UTC & Coordinated Universal Time\\

\end{tabular}
}

\appendixtitles{yes}
\appendixstart
\appendix

\section[\appendixname~\thesection]{Accuracy of the Trajectory Model}
\label{app:speed_model_accuracy}

We validated the accuracy of the trajectory model (\Cref{subsec:route_reconstruction,subsec:speed_model} of the main paper) by comparing all 518{,}577 vessel movements with the positions and times of the $5.9\times 10^7$ cleaned AIS messages associated with these movements. Namely, we~compared the following:
\begin{itemize}
\item The predicted and message positions at the same times as the original messages; 
\item The distance of the simplified route from the original messages (without timing information); and~\item The predicted times at the positions where the simplified routes pass closest to the original AIS message positions.
\end{itemize}

\begin{figure}[H] 
\centering
		\includegraphics[width=0.525\textwidth, trim=0px 0px 0px 15px, clip]{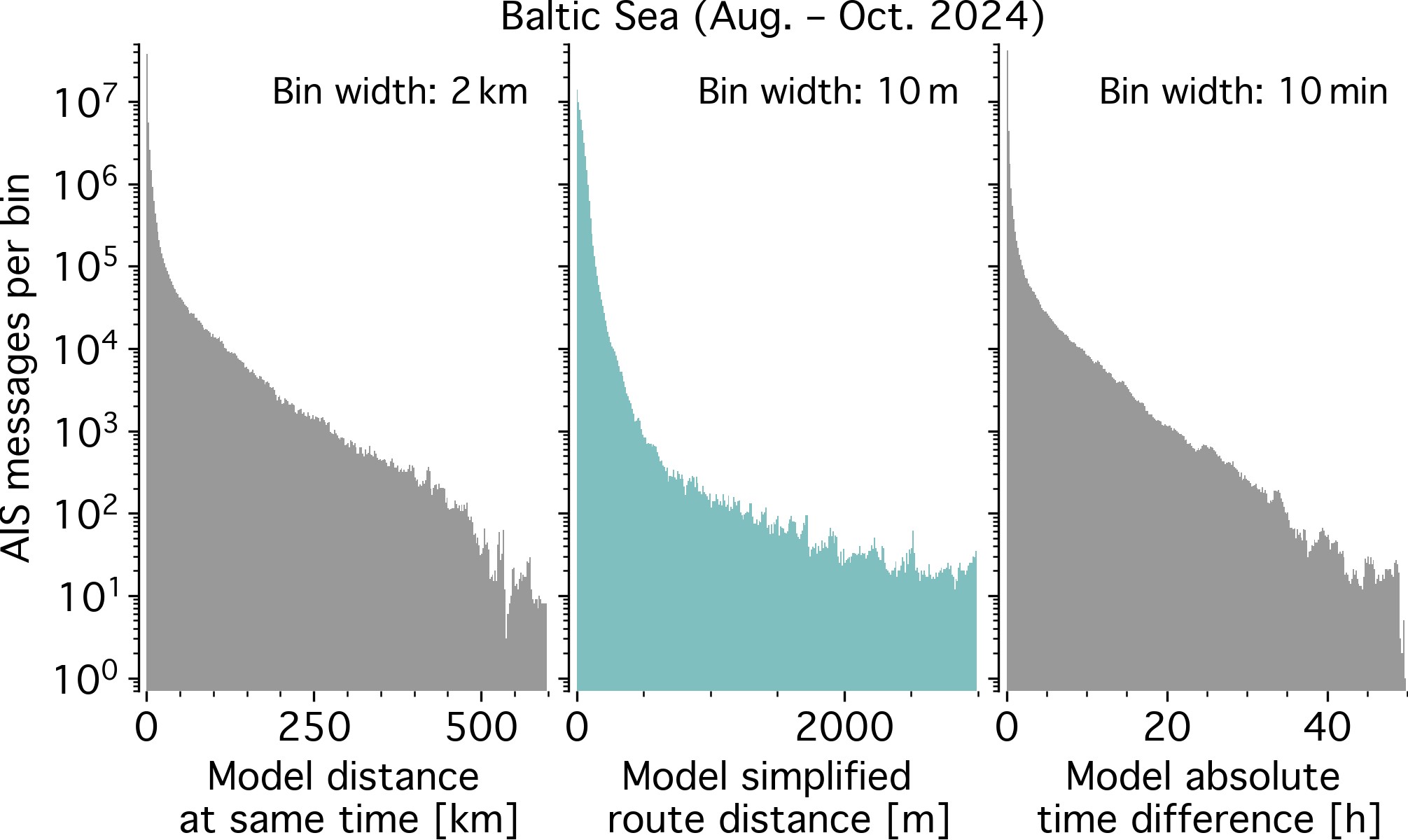} 
	  \caption{Comparison of original AIS message positions and times with the trajectory model. \textbf{Left panel}: Distances between message positions and vessel positions in the model at the corresponding message time. \textbf{Middle panel}: Distances between the message positions and simplified routes. \textbf{Right panel}: Time differences between message times and the times at which the trajectory model has the smallest distance to a message. 
      }
      \label{fig:model_analysis_baltic}
\end{figure}

These comparisons take into account that routes may pass close to the same location multiple times, e.g.,~when a vessel is moving in circles, by~applying an adaptive time window search. \Cref{fig:model_analysis_baltic} presents these comparisons. The~median distance between the trajectories' vessel positions and message positions at reported time is \SI{585}{\meter}, with~90\% of deviations being below \SI{8}{\km} (Figure~\ref{fig:model_analysis_baltic}, left panel). Ignoring timing information, the~middle panel of \Cref{fig:model_analysis_baltic} (teal histogram) shows the impact of the route simplification by the RDP algorithm alone. The~original messages' median distance from the routes is $\SI{23}{m}$, with~3\% of the distances exceeding the $\SI{100}{m}$ threshold. Large outliers occur on long-distance tracklets due to the fact that the RDP algorithm uses Euclidean geometry. The~model’s median time deviation is \SI{112}{\second}, with 90\% of cases being below \SI{25}{\minute} (Figure~\ref{fig:model_analysis_baltic}, right panel). Relatively, these figures translate into a median deviation of vessel positions of 1.2\% of the route length, and~also a median relative timing deviation of 1.2\% of the trajectory~duration.

\begin{figure}[H] 
\centering
		\includegraphics[width=0.5\textwidth, trim=0px 0px 0px 15px, clip]{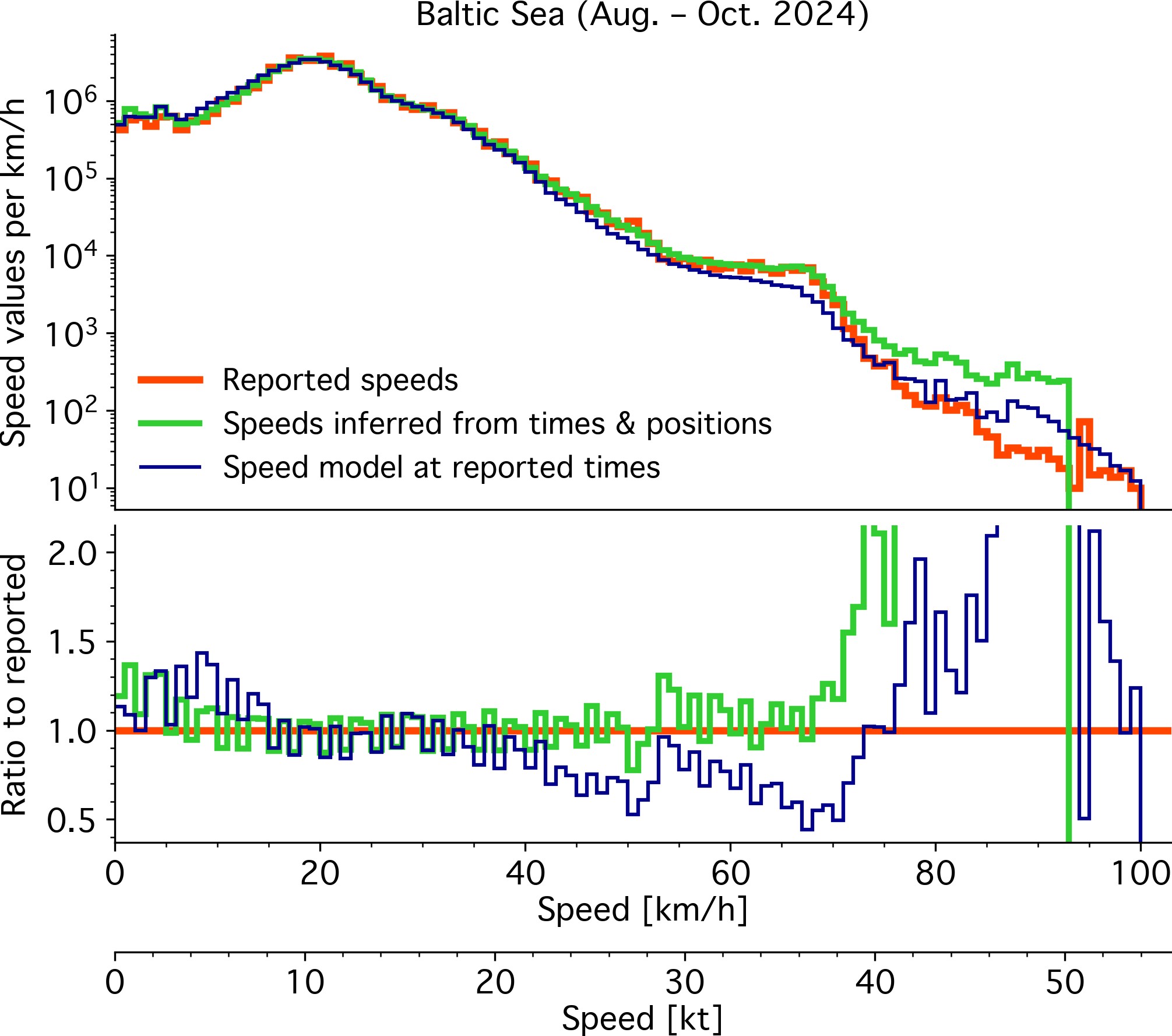} 
	  \caption{
      Vessel speeds from the AIS position-report messages (thick red curves), inferred from reported positions and times (green curves), and~from the trajectory model at the message times (thin blue curves). The~bottom panel shows the relative deviation from the reported values. The~green curve shows the same data as in \Cref{fig:ais_signals_histograms_baltic}c, except~omitting in  
      this figure the static reports. Also, values are grouped in bins of constant width of \SI{1}{\km/\hour} in this figure.
      }
      \label{fig:speed_analysis_baltic}
\end{figure}

We also compared the speed values from all position reports delivered with the AIS messages (thick red curve in \Cref{fig:speed_analysis_baltic}) with the speeds inferred from positions and times used to derive speed control points in the analysis (green curves), and~with the speeds resulting from the trajectory model at the message times (thin blue curves). For  speeds~$\lesssim \SI{20}{\knot}$ (\SI{40}{\km/\hour}), all speed descriptions align well. At~higher speeds, corresponding to about $2\%$ of all messages, the~trajectory model underestimates speeds by up to 50\% and overestimates the highest speeds. This is attributed to the model's inability to capture rapid speed changes and peaks (seen, e.g.,~around 8:00 (mile 36) in the example trajectory of Figure~\ref{fig:speed_model_example}).

\unskip

\section[\appendixname~\thesection]{Mean Segment Length of Lines Crossing a Rectangle Under a Fixed Angle}
\label{app:line_segment_fixedangle}

\begin{figure}[H]
\centering
		\includegraphics[width=0.45\textwidth]{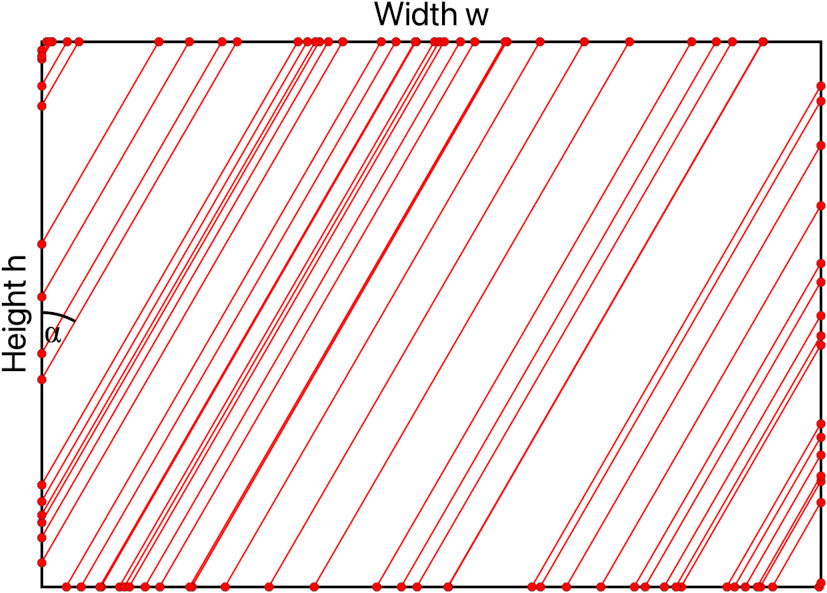} 
			  \caption{\textls[-5]{100 
 straight lines randomly intersecting at $\alpha = 30^\circ$ a rectangle with $r = 10/7$. The~observed mean is $\bar{d}=0.78$ h, and~the expected mean, Equation~(\ref{eq:line_segment_const_angle}), is $\bar{d}({\alpha=30^\circ},\,{r=10/7})=0.82$ h.}} 
   \label{fig:line_segments_angle}
\end{figure}

The mean segment length $\overline{\Delta y}$ through an area $A$ along an axis $y$ is given by 
integration over the orthogonal axis $x$,
\begin{align}
\overline{\Delta y} \,= \,\frac{1}{\Delta x} \int_{x_{\text{min}}}^{x_{\text{max}}} \frac{\mathrm{d} A(x)}{\mathrm{d}x}\, \mathrm{d}x\,,
\label{eq:line_segment_const_angle_abstract}
\end{align}
with $\Delta x = x_{\text{max}}-x_{\text{min}}$, the extent of the rectangle in the coordinate system 
$(x,y)$. For~a rectangle, the~area is always
\begin{align}
A \,= \, \int_{x_{\text{min}}}^{x_{\text{max}}} \frac{\mathrm{d} A(x)}{\mathrm{d}x}\, \mathrm{d}x\, = hw\,,
\end{align}
the product of height $h$ and width $w$, whatever orthogonal coordinates $x$, $y$ are chosen, and~it is
\begin{align}
\overline{\Delta y} \,= \,\frac{hw}{\Delta x} 
\end{align}
in all rotated orthogonal coordinate systems. Rotating the coordinate system by an angle $\alpha$ gives a projected extent $\Delta x' = w\,|\cos\alpha| + h\,|\sin\alpha|$ of a rectangle upright in the system $(x,y)$. Thus, the~average segment length $\overline{\Delta y'}$ in the direction orthogonal to $x'$, with~$\alpha$ measured clockwise from $y$, the~vertical edge $h$, is:
\begin{align}
\overline{\Delta y'} \,= \, \frac{r}{r\,|\cos\alpha| + |\sin\alpha|}\,h\,=:\,\overline{d}({\alpha},\,r)
\label{eq:line_segment_const_angle}
\end{align}
with $r=w/h$. (See also \url{https://math.stackexchange.com/q/3636050} (version 2020-04-21 accessed on 20 November 2025).
\Cref{fig:line_segments_angle} shows an illustration of 100 random lines with $\alpha=30^\circ$, uniformly distributed along $x'$.

\unskip


\section[\appendixname~\thesection]{Supplementary Tables}
\label{app:tables}

\vspace{-6pt}\begin{table}[H]
\centering
\caption{
Definition of transit areas. All areas are defined as rectangular boxes bound by lines of constant latitude and longitude. The~V\"anern Lake transit area is split into two separate boxes for the Nordre \"alv (1) and G\"ota \"alv (2) arms of the Trollh\"atte Canal.
}
\label{tab:exit_areas}
\renewcommand{\arraystretch}{1.2}
\begin{tabularx}{\textwidth}{lCCCC}
\toprule
  \multirow{2}{2.5cm}{\textbf{Name}} & 
   $\boldsymbol{\varphi_\textbf{min}}$ & 
   $\boldsymbol{\varphi_\textbf{max}}$ & 
   $\boldsymbol{\lambda_\textbf{min}}$ & 
   $\boldsymbol{\lambda_\textbf{max}}$ 
\\
 &
  \textbf{ [$\boldsymbol{^\circ}$]} & 
  \textbf{ [$\boldsymbol{^\circ}$]} & 
   \textbf{[$\boldsymbol{^\circ}$]} & 
  \textbf{ [$\boldsymbol{^\circ}$]} 
   \\
\midrule
 Skagerrak (small) & 
 $57.050$ & $58.667$ & $\;\,9.0\;\,\;\,$ & $\;\,9.02\;\,$
 \\
 Skagerrak (default) & 
 $57.050$ & $58.667$ & $\;\,9.0\;\,\;\,$ & $\;\,9.05\;\,$
 \\
 Skagerrak (large) & 
 $57.050$ & $58.667$ & $\;\,9.0\;\,\;\,$ & $\;\;\,9.10\;\,$
 \\[0.1cm]
 Kiel Canal (small)  & 
 $54.3636$    & $54.371$ & $10.140$ & $10.145$  \\
 Kiel Canal (default)  & 
 $54.3636$          & $54.371$ & $10.140$ & $10.150$  \\ 
 Kiel Canal (large)  & 
 $54.3636$    & $54.371$ & $10.140$ & $\;10.160$  
 \\[0.1cm]
 Limfjord &  
 $56.535$ & $57.050$ & $\;\,9.0\;\,\;\,$ & $9.05$ \\
Oder River &  
 $53.343$ & $53.385$ & $14.493$ & $14.621$\\
Telemark &  
 $59.097$ & $59.130$ & $\;\,9.480$ & $\;\,9.747$\\
V\"anern Lake 1 &  
 $57.766$ & $57.806$ & $11.805$ & $11.905$\\
V\"anern Lake 2 &  
 $57.677$ & $57.719$ & $11.902$ & $12.002$\\
V\"attern Lake &  
 $58.384$ & $58.476$ & $16.620$ & $16.680$\\
S\"odert\"alje& 
 $59.163$ & $59.200$ & $17.631$ & $17.708$\\
Stockholm &  
 $59.2918$    & $59.459$ & $18.025$ & $18.084$\\
Saimaa Canal & 
$60.700$ & $60.730$ & $28.619$ & $28.830$\\
Neva River &  
 $59.857$ & $60.008$ & $30.259$ & $30.312$
 \\
\bottomrule
\end{tabularx}
\end{table}
\unskip

\begin{table}[H]
\centering
\caption{Mapping 
 of vessel type codes in the AIS-A ship static messages to the categories in which the results are presented in this work. No codes above 100 were observed in the analysis~period.}
\label{tab:vessel_types}
\newcolumntype{C}{>{\centering\arraybackslash}X}
\begin{tabularx}{\textwidth}{LL}
\toprule
 \textbf{Vessel Type Category} & \textbf{Vessel Type Codes} \\
\midrule
 Passenger, high-speed & 
 20, 23--29, 40--49, 60--69
 \\[0.15cm]
 Law enforcement, military & 
35, 55
 \\[0.15cm]
 Cargo & 
 70--79
 \\[0.15cm]
 Pilot, tug, rescue, diving/dredging  & 
21, 22, 31--34, 50--58
  \\[0.15cm]
 Tanker & 
80--89
 \\[0.15cm]
 Others, including fishing& 
 all other codes or no code
 \\
\bottomrule
\end{tabularx}
\end{table}
\unskip

\begin{table}[H]
\centering
\caption{
Average number of vessels present in the ROI ($N$) and total inbound and outbound traffic through all transit areas, $\dot{N}$, for~the default (\textit{df}) parameters and variations \textit{hi} and \textit{low} to suppress false-positive and false-negative transit events.
The bottom rows show the relative difference to the default case.
}\label{tab:case_comparison}
\begin{tabularx}{\textwidth}{CCC}
\toprule
 & \textbf{Vessels in ROI, $\boldsymbol{N}$} & \textbf{Transits/Day, $\boldsymbol{\dot{N}}$} \\
\midrule
Case \textit{low} & $4038.9$ & $\;\,347.1$ \\
Default (case \textit{df})   & $4061.1$ & $\;\,313.1$ \\
Case \textit{hi}   & $4256.2$ & $\;\,214.6$ \\\midrule
$\delta_\text{low}$ & $-0.5\%$ & $+\,10\%$ \\
$\delta_\text{hi}$  & $+4.8\%$ & $-\,31\%$ \\
\bottomrule
\end{tabularx}
\end{table}
\unskip

\begin{table}[H]
\centering
\caption{Fraction 
 for AIS-B vessels, $\delta_\text{ais-b}$, among~all vessels in the Baltic Sea. The~percentages for the 
 first five
 categories 
  are estimated from the October 2024 Baltic Sea AIS data. The
  percentages 
  in the vessel categories ``Others including fishing'' and GT $<10{,}000$
 are normalized to the yearly total average using the category results from \Cref{tab:vessel_numbers,tab:vessel_numbers_gt}.}
\label{tab:vessel_aisb}
\renewcommand{\arraystretch}{1.2}
\begin{tabularx}{\textwidth}{LC}
\toprule 
\textbf{Vessel Type} &  \textbf{$\delta_\text{ais-b}$} 
  \\
\midrule
 All vessels (assumed yearly average)& 
  $30\%$  
  \\\midrule
 Passenger, high-speed & 
 $11\%$
 \\
 Law enforcement,  military & 
 $20\%$ 
 \\
 Cargo  & 
 $13\%$ 
 \\
 Pilot, tug, rescue, diving/dredging  & 
 $18\%$
  \\
 Tanker  & 
 <$1\%$ 
 \\[0.1cm]
 Others including fishing (derived)&   $55\%$
 \\\midrule
 GT $<10{,}000$ (derived)& $37\%$ \\ 
 GT $\geq10{,}000$ (assumed) & $\;\,0\%$
 \\
\bottomrule
\end{tabularx}
\end{table}

\isPreprints{}{
\begin{adjustwidth}{-\extralength}{0cm}
} 

\reftitle{References}

\isPreprints{}{
\end{adjustwidth}
} 
\end{document}